\documentclass[oldversion]{aa}  
\usepackage{graphicx}
\usepackage{epsfig}

\usepackage{txfonts}
\usepackage{natbib}
\bibpunct{(}{)}{;}{a}{}{,} 
\begin{document}
   \title{On the frequency, intensity and duration of starburst episodes \\ triggered by galaxy interactions and mergers}

   \author{P. Di Matteo\inst{1,2}
	\and
	F. Bournaud\inst{3}
        \and
        M. Martig\inst{3}
	\and
	F. Combes\inst{1}
        \and
        A.-L. Melchior\inst{1,4}
        \and
        B. Semelin\inst{1,4}
        }


   \offprints{P. Di Matteo}

   \institute{Observatoire de Paris, LERMA, 61, Avenue de L'Observatoire, 75014 Paris, France  \and  Observatoire de Paris, Section de Meudon, GEPI,  5 Place Jules Jannsen, 92195, Meudon, France \and
   Laboratoire AIM, CEA-Saclay/DSM/IRFU/SAp -- CNRS -- Universit\'e Paris Diderot, 91191 Gif-sur-Yvette, France \and Universit\'e Pierre et Marie Curie - Paris 6, 4 Place Jussieu, 75252 Paris Cedex 5, France}

   \date{Received ; Accepted }

\abstract{We investigate the intensity enhancement and the duration of starburst episodes, triggered by major galaxy interactions and mergers. To this aim, we analyze two large statistical datasets of numerical simulations. These have been obtained using two independent and different numerical techniques to model baryonic and dark matter evolution, that are extensively compared for the first time. One is a Tree-SPH code, the other one is a grid-based N-body sticky-particles code. We show that, at low redshift, galaxy interactions and mergers in general trigger only moderate star formation enhancements. Strong starbursts where the star formation rate is increased by a factor larger than 5 are rare and found only in about 15\% of major galaxy interactions and mergers. Merger-driven starbursts are also rather short-lived, with a typical duration of the activity of a few $10^8$~yr. These conclusions are found to be robust, independent from the numerical techniques and star formation models. At higher redshifts where galaxies contain more gas, gas inflow-induced starbursts are neither stronger neither longer than their local counterparts. In turn, the formation of massive gas clumps, results of local Jeans instability that can occur spontaneously in gas-rich disks or be indirectly favored by galaxy interactions, could play a more important role in determining the duration and intensity of star formation episodes.} 
 

  \maketitle

%

\section{Introduction}

The role played by galaxy interactions in affecting star formation was realized by \citet{larson}, who showed that disturbed galaxies in the Arp Catalogue \citep{arp} have a larger dispersion in their colors and a bluer envelope in the ($U$-$B$, $B$-$V$) plane than normal systems taken from the Hubble Atlas \citep{sandage}. Using evolutionary synthesis models, they suggested that the features found in $UBV$ colors of interacting galaxies were caused by bursts of star formation lasting a few $10^7-10^8$ years. The large amount of observational works that followed (see \citet{saas} for a complete review) showed that in the Local Universe, the response of galaxies to mutual interactions and mergers is quite varied.

Many starbursts in the Local Universe take place in the central regions of interacting/merging galaxies, as it is the case for instance for  NGC~7714 studied by \citet{weed81}, and the protoype starburst galaxy M~82 \citep{deg01a,deg01b}. Another well-studied example is the NGC~4038/4039 system (the Antennae): this early stage merger presents an extended star formation, the most intense star forming regions being located between the two galaxies \citep{wang04}. 
Actually, the vast majority of UltraLuminous Infrared Galaxies (ULIRGs) at low redshift, i.e. the strongest starbursts in the Local Universe, are found in interacting and merging galaxies (\citet{sanmir96}, see also \citet{ducmm97}). This is however not reciprocal. Indeed, \citet{berg03} among others have shown that, in a magnitude-limited sample of 59 interacting and merging galaxies, only a weak enhancement of star formation (a factor of 2-3 in the galaxy centers) is found when compared to a reference sample of non-interacting galaxies, so that the contribution of interactions and mergers to the global star formation activity at low redshift could on average be much less efficient than suggested by the strongest examples of starbursts.

At high redshift, the role of mergers in the star formation history of galaxies is debated, too. In a pioneering study of distant infrared galaxies, \citet{elbces03} showed that the majority of present-day stars were formed in dusty starbursts, and suggested that the later were triggered by galaxy interactions. \citet{cons03} also suggested that about two thirds of submillimeter galaxies at $z>1$ are undergoing a major merger. In a study of the Spitzer First Look Survey, \citet{bridge07} argued that close pairs are major contributors to the star formation density at $z>0.7$, about half of the star formation rate density at $z \sim 1$ being attributed to major mergers and interactions. However, an important part of the infrared luminosity of galaxies could be caused by AGN heating  \citep[e.g.][]{daddi07a, daddi07b}. \citet{bell05} find that less than one third of actively star forming galaxies at $z \sim 0.8$ are actually interacting or merging, the majority having undisturbed disk morphologies. Similarly, selecting interacting galaxies in the GEMS survey, \citet{jogee07, jogee08} find that over the redshift interval z$\sim$0.24 to 0.80 (corresponding to   lookback times 3 to 7 Gyr), the average   SFR of strongly distorted interacting/merging massive galaxies is only
 modestly  enhanced with respect to  normal undisturbed galaxies. At even higher redshift ($z \sim 2$) in GOODS, \citet{daddi07a}  find that the star formation activity of ULIRGS is long lived (at least half a Gyr) which might be longer than expected for merger-induced starbursts.

In the last decade, a lot of progress has been made in our theoretical understanding of the role played by galaxy interactions in driving star formation, and a large variety of results has been obtained (see \citet{stru05} for a review). Barnes \& Hernquist (1991) have shown that tidal interactions between galaxies can drive gas inflows towards the central regions, which can increase the star formation rate. \citet{mih94a, mih96} directly studied the star formation activity in mergers of equal mass disk galaxies using self-consistent N-body simulations \citep{mih94b} including stars, dark matter, and gas dynamics. They also pointed out that the timing and strength of an interaction-driven starburst depend on the morphology of the interacting systems, bulgeless disk galaxies being more prone to suffer enhanced star formation in the first phases of the interaction, rather than in the merging phase. Important episodes of star formation can also arise in the case of minor merger events \citep{mih94c}, but major mergers of galaxies of comparable masses are the most efficient situations to trigger strong starbursts. Indeed, \citet{cox07} have shown that the star formation activity of merging galaxies decreases rapidly with increasing mass ratios.
More sophisticated models including supernovae feedback in regulating star formation have also been explored \citep{spri00, cox06, cox07}. In particular, \citet{cox06} pointed out that the large amount of freedom in selecting the feedback parameters could play a significant role in determining the maximum star formation rate during a galaxy merger. Indeed, supernovae feedback regulates the star formation and, on average, reduces the intensity of merger-induced starbursts. The effect of feedback is however generally modest within the assumptions considered as the most realistic ones.

Nevertheless, that some cases of major interactions or mergers can trigger strong starbursts, as in the examples shown by \citet{mih96}, does not imply that the enhancement of the star formation activity is systematically high. A sample of about 50 Nbody-SPH simulations of galaxy interactions was performed by \citet{kap05}, and these authors found  that the integrated star formation rate during an interaction is moderately increased, up to a factor of 5 but on average a factor of 2 with respect to that of isolated galaxies. More recently, \citet{dimatteo07} (hereafter DM07) presented more than two hundreds simulations  of galaxy interactions and mergers, restricted to coplanar cases, pointing out the difficulty to drive intense starbursts. Only 17\% of the mergers in their sample have strong bursts with an SFR ten times higher than in isolated galaxies, and half of the sample shows  an enhancement of the SFR by a factor no bigger than 4 at the peak of the starburst.

In this paper, we extend the analysis started in DM07 on the relation between major galaxy interactions and star formation. On the one side, we enlarge the sample of Tree-SPH simulations studied in that paper, modeling orbits with various inclinations. On the other side, we test whether the main conclusions depend on the model used. To this aim, we compare the result of this first sample to a second (smaller) sample of simulations performed with a different numerical code -- a particle-mesh sticky-particle  (hereafter PM-SP) -- using somewhat different initial conditions and testing different numerical recipes for star formation. This large and heterogeneous data set shoud help in finding robust results about the starbursts--galaxy interaction connection. It is also, to our knowledge, the first numerical work in this field where simulations realized with different codes and star formation recipes are directly compared.

The layout of the paper is as follows. The simulations techniques and parameters are presented in  Sect.\ref{description}, distinguishing Tree-SPH and PM-SP models. Results from Tree-SPH simulations are presented in Sect.\ref{treeresults}, and, in Sect.\ref{comparison} the main findings are compared to that obtained employing a PM-SP approach. After comparing  our results about the frequency of starburts episodes with observations (Sect.\ref{obs}), in Sect.\ref{concl}, the main conclusions of this research are presented. 


\section{Description of the numerical simulations}\label{description}

\subsection{The data set}\label{ini}

In DM07, a statistical study of the relationship  between star formation and galaxy interactions was performed using a  set of 216 simulations. The morphology of the interacting galaxies,  as well as their encounter velocity and distance, were varied,  and the role played by different parameters (gas fraction, galaxy  minimum separation, galaxy relative velocity, strength of tidal  effects, etc) was analyzed. That work represented also the first  numerical work, to our knowledge, where  the  difficulty to drive substantial bursts of star formation during major  galaxy encounters was outlined. The results presented in that paper showed,  indeed, that major mergers were neither always responsible for triggering  intense bursts of star formation, nor for converting large quantities  of gas mass into new stars. However, the influence of the orbit  inclination was not studied, all disk galaxies in this first sample being coplanar.  Also, the influence of the numerical techniques and  recipes to compute the star formation rate has not been studied. The  goal of the present work is to remove these limitations.

First, we extend the analysis of the relation between galaxy  interactions and star formation to a larger parameter set, taking  into account also encounters between galaxies having not zero disk inclinations. Indeed, coplanar interactions are quite peculiar cases that can favor the formation of regions of gas shocks, thus  it is interesting to explore to what extent removing this assumption can modify the star formation evolution of the pair. We will also study the star formation  activity when the two interacting galaxies have an initial amount of  gas higher than that of galaxies in the Local Universe. This initial work is here extended with the realisation of a sample of 672 simulations (648 for local galaxies and 24 for high-redshift systems) made with the same Tree- SPH code as in DM07, whose main characteristics are  recalled below.

To study whether the main conclusions depend on the numerical  techniques and star formation schemes adopted, we have run a second  set of 96 simulations with a different numerical code (a particle-mesh  code with a sticky-particle modeling of the ISM),
 where the initial conditions are also somewhat  different. Some simulations in this sample also employ star formation  models that differ from the Schmidt law. This is in order to understand  how and in which way different star formation recipes can affect the  star formation evolution during galaxy interactions, and if the  conclusions of the main dataset are crucially dependent on the star  formation model or not.

In our opinion, the study of such a large and vast data set should  help in finding strong and robust results about the star formation  efficiency during galaxy major interactions, avoiding some of the  limitations which have affected previous numerical works.

\subsection{Numerical methods}\label{numerical}
\subsubsection{Tree-SPH simulations}\label{numerical1}
The first set of 672 simulations has been realized employing the same  Tree-SPH code as in DM07. Gravitational forces are calculated using a  hierarchical tree method \citep{bh86} and gas evolution is followed  by means of smoothed particle hydrodynamics \citep[SPH,][] {lucy77,gm82}. The code and the adopted numerical parameters are  described in DM07 and references therein; we here recall only the  main features. Gravitational forces are calculated using a tolerance  parameter $\theta=0.7$ and include terms up to the quadrupole order  in the multiple expansion. A Plummer potential is used to soften  gravity at small scales, with constant softening lengths  $ \epsilon=280\ \mathrm{pc}$ for all species of particles. The  equations of motion are integrated using a leapfrog algorithm with a  fixed time step of 0.5~Myr. A conventional form of the artificial viscosity is used in the SPH  model, with parameters $\alpha=0.5$ and $\beta=1.0$ \citep{hk89}. To  describe different spatial dynamical ranges, SPH particles have  individual smoothing lengths $h_i$, calculated in such a way that a  constant number of neighbors lies within  $2h_i$. The simulations in  this paper have been performed using a number of neighbors $N_s\sim 15 $. The interstellar gas is modeled as isothermal, with a temperature  $T_{gas}=10^4 K$.\\
In  \citet{benoit02}, a standard validation test for this code (the collapse of an initially static, isothermal sphere of self-gravitating gas) has  been presented. No change is made in the algorithm with respect to that paper, and we here use an isothermal equation of state for the gas instead of a multiphase model. Other tests on the dependency of the star formation rates on the numerical parameters adopted are presented in Appendix  \ref{tests}.

\subsubsection{PM-SP simulations}\label{numerical2}

The second set of simulations was performed with a PM-SP ({\it  particle-mesh -- sticky-particles}) code, which is a grid-based N- body code described in \citet{BC03}. The density of particles is  computed on Cartesian grids of maximal resolution 330~pc through a Cloud-In Cell multi-linear interpolation. An FFT technique \citep{james77} is  then used to compute the associated gravitational potential, with a  softening length $\epsilon = 330$~pc.

The collisional dynamics of the ISM is modeled with a sticky- particle scheme: gas particles, that model interstellar gas clouds,  undergo inelastic collisions during which their relative velocity is  multiplied by $\beta_r$ along the direction of their positions, and $ \beta_t$ in the perpendicular direction. Here, we use $\beta_r = 0.7$  and $\beta_t = 0.5$, which creates a cool enough medium for thin  spiral arms to form in isolated disks but without making the disk  unstable to axisymmetrical perturbations. In the case of particularly gas-rich galaxies (see Sect.~\ref{comp_gas}), the dissipation is reduced, using $\beta_r = 0.8$  and $\beta_t = 0.7$, in order to limit the formation of clump instabilities.

\subsection{Star formation recipes}\label{recipe}
Different recipes and numerical methods have been adopted in existing models to include star formation and account for the effects that this star formation has on the surroundings \citep {kat92,ste94,spri00,spr03,cox06}. These are usually based on the so-called Schmidt-Kennicutt law (Schmidt 1959, Kennicutt 1998a,b) and assume that the local star formation rate can be inferred from the local gas density, sometimes combined with a stability threshold. In the following two subsections, the different numerical recipes adopted in our codes are described. Their main features are summarized in Table \ref{SFtable}.

   \begin{table}
      \caption[]{Main features of the star formation recipes adopted  in the Tree-SPH code and in the grid-sticky particles one.}
         \label{SFtable}
     \centering
      \begin{tabular}{lccc}
        \hline\hline
        & & Tree-SPH & PM-SP\\
        \hline
        & $\Sigma_{SFR}\propto {\Sigma_{gas}}^{1.5}$ & yes & yes\\
        SF laws & $\Sigma_{SFR}\propto {\Sigma_{gas}}\Omega$  & no & yes\\
        & $\Sigma_{SFR}\propto {\Sigma_{gas}}^{1.5}\Omega$ & no & yes\\
        & & & \\
        SF feedback & kinetic feedback & yes & no\\
        & metal enrichment & yes & no\\
            \hline

            \hline
         \end{tabular}
   \end{table}

\begin{figure*}
  \centering
  \includegraphics[width=6cm,angle=0]{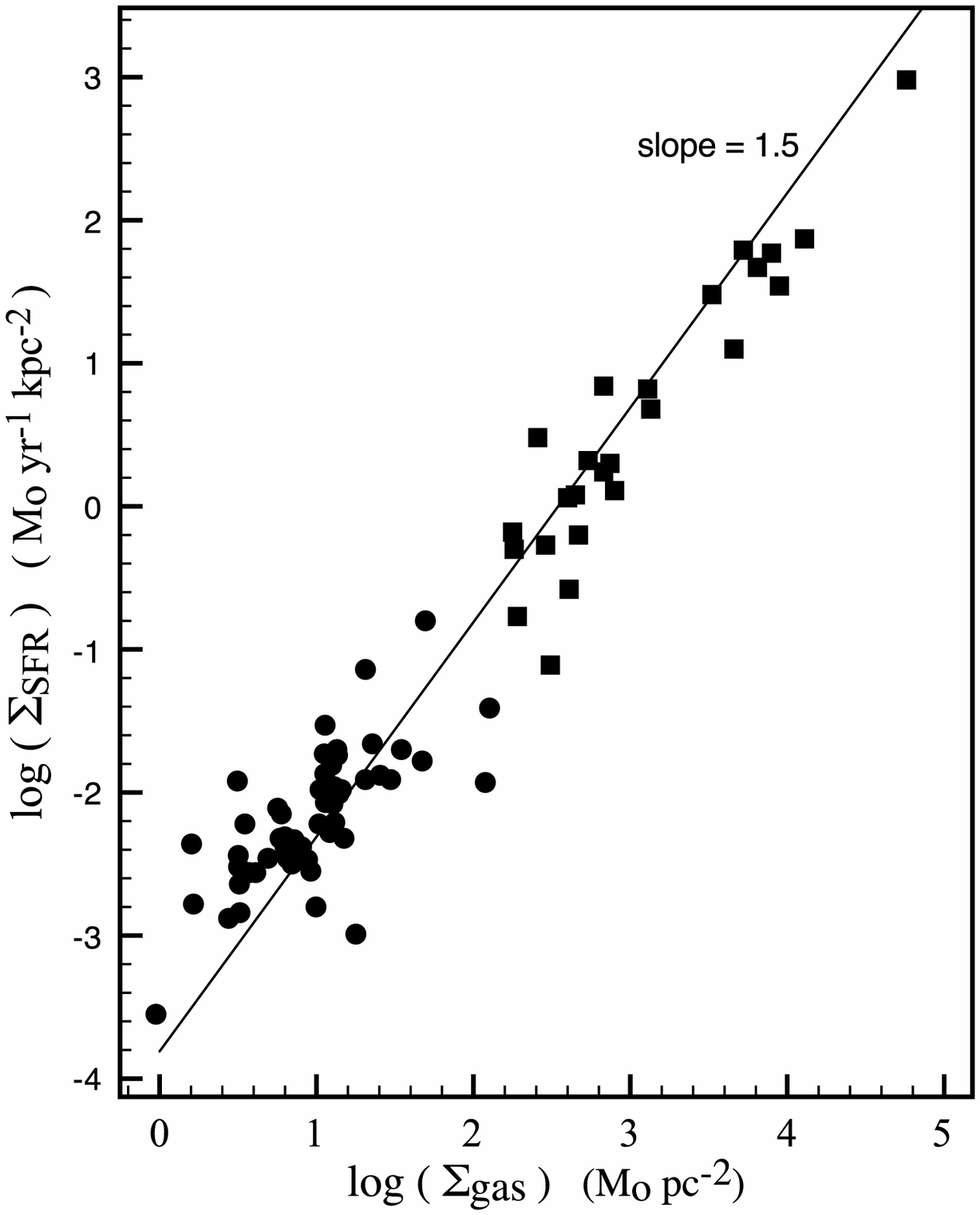}
  \includegraphics[width=6cm,angle=0]{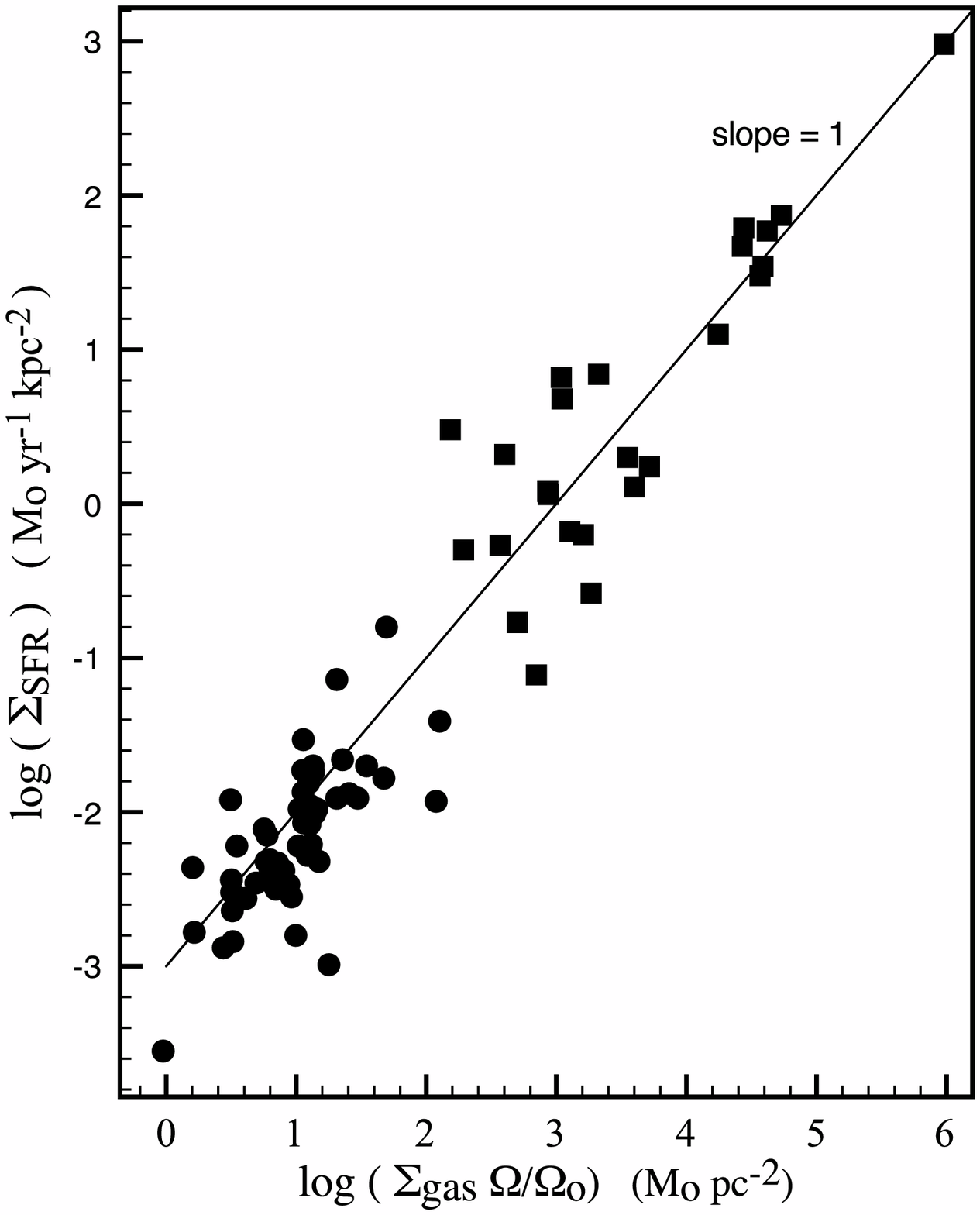}
  \includegraphics[width=6cm,angle=0]{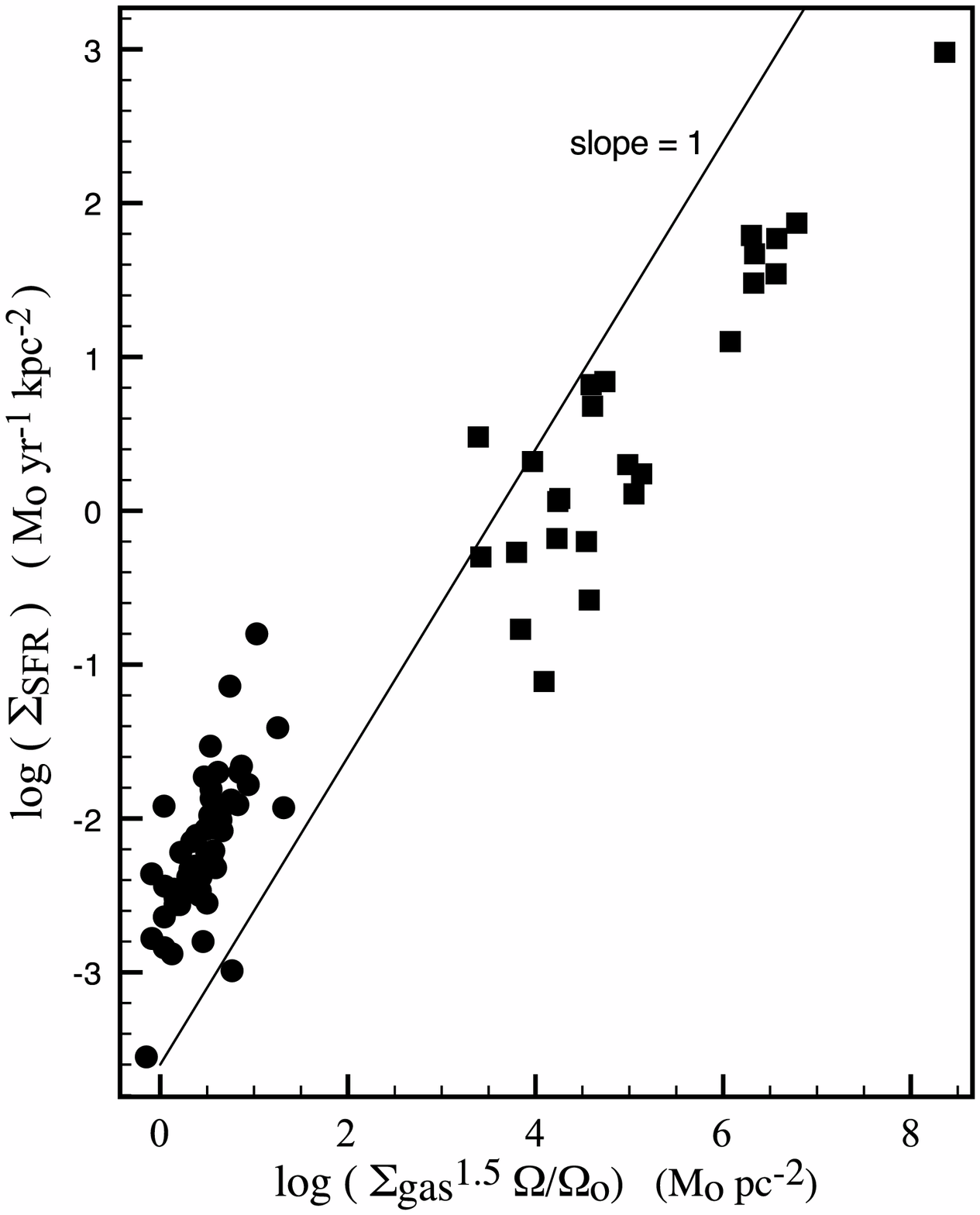}
\caption{Comparison of the three star formation models used in the PM-SP simulations to observational data from \citet{ken98b}. Left:  Schmidt law with a 1.5 exponent -- Middle: $\Sigma_{SFR} \propto  \Sigma_ {gas} \times \Omega$ model, also compatible with observations  -- Right: $\Sigma_{SFR} \propto \Sigma_ {gas}^{1.5} \times \Omega$  model. This last model is barely compatible with observations and  tends to overestimate the star formation rate at high gas density, we  thus use this model to put an upper limit on the actual starburst  efficiencies.}\label{sfrobs} 
\end{figure*}

   \begin{table*}
      \caption[]{Tree-SPH and PM-SP simulations: Galaxy parameters.  The bulge and the halo are modeled as Plummer spheres, with  characteristic masses $M_B$ and $M_H$ and characteristic radii $r_B$  and $r_H$.  $M_{*}$ and  $M_{g}$ represent the masses of the stellar  and gaseous disks, whose vertical and radial scale lengths are given,  respectively, by $h_{*}$ and $a_{*}$, and $h_{g}$ and $a_{g}$.}
         \label{galpar}
     \centering

                     \begin{tabular}{llccccccccc}
            \hline\hline
            && \multicolumn{5}{c}{Tree-SPH} && \multicolumn{3}{c}{PM- SP} \\
        && gE0 & gSa & gSb & gSd & gSb+ && gSb0 & gSb & gSb+\\
            \hline
        $M_{B}\ [2.3\times 10^9 M_{\odot}]$ && 70 & 10 & 5 & 0 &  5        && 5    &  5   & 5    \\
        $M_{H}\ [2.3\times 10^9 M_{\odot}]$ && 30 & 50 & 75 & 75 & 75   && 65  &  65 & 65  \\
        $M_{*}\ [2.3\times 10^9 M_{\odot}]$ && 0 & 40 & 20 & 25 & 20      && 22  &  20 &  12 \\
        $M_{g}/M_{*}$ && 0 & 0.1 & 0.2 & 0.3 &  0.5                                  && 0    & 0.15& 0.50 \\
        & & & & &&& & && \\
        $r_{B}\ [\mathrm{kpc}]$ && 4 & 2 & 1& -- & 1&& 1.8 & 1.8 & 1.8  \\
        $r_{H}\ [\mathrm{kpc}]$ && 7 & 10 & 12 & 15 & 12&& 10. & 10. & 10.\\
        $a_{*}\ [\mathrm{kpc}]$ && -- & 4 & 5 & 6 & 5    && 5. &5.& 5.\\
        $h_{*}\ [\mathrm{kpc}]$ && -- & 0.5 & 0.5 & 0.5 & 0.5&&  0.7& 0.7& 0.7 \\
        $a_{g}\ [\mathrm{kpc}]$ && -- & 5 & 6 & 7 & 6&& 15.& 15.& 15. \\
        $h_{g}\ [\mathrm{kpc}]$ && -- & 0.2 & 0.2 & 0.2 & 0.2&&  0.25& 0.25& 0.25 \\
        & & & & &&& & && \\
	$Q_{gas}$ && -- & 0.8 & 0.8 & 0.8 & 1. && 1.2 &1.2 &1.2\\
            \hline
         \end{tabular}
   \end{table*}

   \begin{table*}
         \caption[]{Tree-SPH and PM-SP simulations: Particle numbers  for each galactic component}
         \label{numbers}
     \centering
         \begin{tabular}{llccccccc}
            \hline\hline
            && \multicolumn{4}{c}{Tree-SPH} && \multicolumn{2}{c}{PM- SP} \\
        && gE0 & gSa & gSb, gSb+ & gSd && gSb0 & gSb,gSb+ \\
            \hline
        $N_{gas}$ && -- & 20000 & 40000 & 60000 && -- & 100000 \\
        $N_{star}$ && 80000 & 60000 & 40000 & 20000 && 70000 & 70000 \\
        $N_{DM}$ &&40000 & 40000 & 40000 & 40000 &&  50000 & 50000 \\
            \hline
         \end{tabular}
   \end{table*}

\subsubsection{Tree-SPH simulations}
In all the simulations performed adopting the Tree-SPH code, a  density-dependent star formation law (Schmidt law) has been employed.  The star formation rate is locally given by:

\begin{equation}\label{loc}
\rho_{SFR}=C \times {\rho_{gas}}^{1.5}
\end{equation}
with the constant $C$ chosen such that the isolated disk galaxies  form stars at an average rate of between  1 and 2.5~M$_{\odot}$~yr$^{-1}$. This parametrization is consistent with the observational  evidence that on global scales the SFR in disk galaxies is well represented by a Schmidt law with an exponent 1.4 \citetext{see \citealp{ken98a,ken98b}, but also \citealp{wb02,bois03,gs04}}.

Once the SFR recipe is defined, we apply it to SPH particles, using  the ''hybrid'' method described in \citet{mih94b}. This consists in representing each gas particle with two mass values, one  corresponding to its gravitational mass $M_i$, whose value stays  unchanged during the whole simulation, and the other describing the  gas content of the particle $M_{i,gas}$, whose value changes over  time, according to Eq.\ref{loc}. Gravitational forces are always evaluated on the gravitational mass $M_i$, while hydrodynamical quantities, in turn, uses the time-varying mass of gas $M_{i,gas}$. If the gas fraction present in the hybrid particles drops below  $5\%$ of the initial gas content, the hybrid particle is totally converted  into a star-like particle and the small amount of gas material still present is spread over neighbors.

We also followed the method described in \citet{mih94b} for including  the effects of star formation into the interstellar medium (metal  enrichment and energy injection in the ISM by supernovae explosions).  The method is fully described by these authors, so we refer  the reader to it for more details.

\subsubsection{PM-SP simulations}

The Tree-SPH simulations have been carried out for a large set of  morphological and orbital parameters (see Sect.~\ref{tsphcond}), but only with a  Schmidt law to compute the star formation rate. The set of orbital  parameter in the PM-SP sample is less extensive (Sect.~\ref{pmspcond}), which enabled us to perform simulations with different models to compute the star formation rate:
\begin{itemize}
\item First, we used the same Schmidt law as in Tree-SPH simulations with an  exponent 1.5, as described in Eq.~\ref{loc}. There is, however, a  difference in the way this prescription is applied. In Tree-SPH simulations, the gas density $ \rho_{gas}$ is the SPH density, computed with an adaptative  resolution. Some theory predicts that the Schmidt law is scale-free  \citep[e.g.,][]{elmegreen02} but this is not necessarily the case. In the PM-SP simulations, we  then compute the density $\rho_{gas}$ on the Cartesian grid with a fixed resolution. The corresponding $\rho_ {SFR}$ indicates a number of gas particles to be converted into  stellar particles in each cell.
\item Second, we model a star formation rate $\rho_{SFR} \propto \rho_ {gas} \times \Omega$, where $\Omega$ is the local rotational angular velocity of  the gas disk. This can model the smaller size and shorter collapse  timescale of molecular clouds at small radii \citep{elmegreen97} and/or their higher collision frequency \citep{silk97}. Just like the  Schmidt law, this is compatible with observations according to \citet{ken98b} -- see also Fig.~\ref{sfrobs}. It is here practical that only one galaxy contains gas in the PM-SP simulations (see Sect.\ref{pmspcond}), so that the time-dependent rotation curve $\Omega(r)$ can be computed for this galaxy. To this aim, we compute the velocity of gas particles in 50~pc wide radial bins every 75~Myr (assuming $\Omega$ is constant in the central 100~pc to avoid singularities). We use the result as a proxy for the actual rotation curve, which could anyway not be clearly defined during the merger. This is a simple way to estimate the influence of the dynamical timescale on the star formation activity during the merger. We checked that using the initial rotation curve of the galaxy instead does not lead to large changes:  the rotation curve evolves mainly during the very final stages of the merger relaxation, while most star formation occurs earlier, so that the way $\Omega(r)$ is estimated is not crucial.
\item Third, we used a model with $\rho_{SFR} \propto \rho_{gas}^ {1.5} \times \Omega$, which is the combination of the two previous  models. This scheme has no theoretical motivation but corresponds to  an \emph{upper limit} of the observed non-linearity of the gas density --  star formation rate density relation (see Figure~\ref{sfrobs}). This model  should thus correspond to an upper limit to the star formation  efficiency in galaxy mergers.
\end{itemize}

These three star formation recipes are applied in 3-D models, but are equivalent to 2-D laws based on the gas surface density $\Sigma$ and SFR surface density, under the reasonable assumption of a uniform thickness of gas disks. Each star formation model is compared to observational data from \citet{ken98b} in Fig.~\ref{sfrobs}.

In the PM-SP simulations, we do not use hybrid particles like in the Tree-SPH model, but  instead convert gas particles into stellar particles at a rate that,  in each cell of the grid, is given by the chosen star formation  model. Not using hybrid particles reduces the mass resolution in the  treatment of star formation, but prevents a newly formed stellar mass  to follow an SPH dynamics. Each method thus has drawbacks, and the two  different assumptions are tested in our two datasets.

Energy feedback from supernovae is not included in the PM-SP  simulations. The reasons for this choice are: {\it (i)} this provides  different assumptions compared to the Tree-SPH sample that includes  feedback; {\it (ii)} within the most standard assumptions, feedback  does not have a major influence on the relative SFR evolution in mergers \citep {cox06} and {\it (iii)} the general effect is to regulate star formation and reduce the starburst efficiency in the merging phase-- because our  conclusion is that most merger-driven starbursts have a low  efficiency, not including feedback is a rather  conservative choice. 

\subsection{Initial conditions}\label{conditions}
\subsubsection{Tree-SPH simulations}\label{tsphcond}
\paragraph{Galaxy models: moving along the Hubble sequence\\}\label{galmod1}

\begin{figure}
  \centering
  \includegraphics[width=8cm,angle=0]{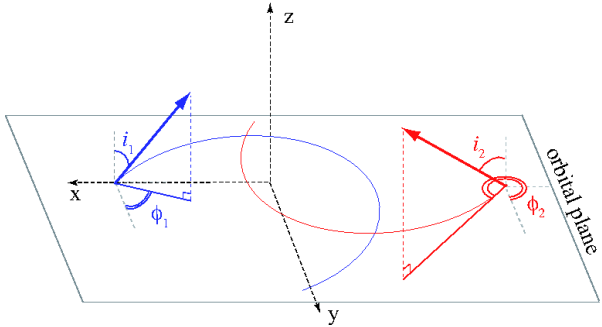}
\caption{Adopted orbital geometry for our simulations. We set up the collision in such a way that the orbital angular momentum is parallel to the $z-$axis and that the centers of the two galaxies are on the $x-$axis initially. The galaxy spins are represented by the blue and  red arrows, respectively. They are specified in terms of the spherical coordinates ($i_1,\Phi_1$) and ($i_2,\Phi_2$). See Table \ref{angles} for their initial values.\label{sketch}}
\end{figure}

   \begin{table}
         \caption[]{Orientation of the galaxy spins, for Tree-SPH and PM-SP simulations.}
         \label{angles}
     \centering
         \begin{tabular}{lcc}
            \hline\hline
	    & Tree-SPH & PM-SP\\
	    \hline
	    $i_1$ & 0$^\circ$ & 33$^\circ$\\
	    $\Phi_1$ & 0$^\circ$ & 30$^\circ$\\
	    $i_2$ & 0$^\circ$,45$^\circ$,75$^\circ$,90$^\circ$ & 0$^\circ$\\
	    $\Phi_2$ & 0$^\circ$ & 0$^\circ$\\
            \hline
         \end{tabular}
   \end{table}

As in DM07, our aim is to exploit a large set of interactions,  involving \emph{galaxies of all morphologies from ellipticals to late- type spirals}.  For each galaxy type, the halo and the bulge (if any)  are modeled as a Plummer sphere \citep[][pag.42]{bt1}, with  characteristic masses $M_B$ and $M_H$ and characteristic radii $r_B$  and $r_H$. The stellar and gaseous disks follow a Miyamoto-Nagai density profile  \citep[][pag.44]{bt1} with masses $M_{*}$ and  $M_{g}$ and  vertical  and radial scale lengths given, respectively, by $h_{*}$ and $a_{*}$,  and $h_{g}$ and $a_{g}$.\\
For the different morphologies adopted, stellar disks have a peak rotation speed within  $a_{*}$ and $3a_{*}$ which is between $63\%$ (for Sbc spirals) and  $75\%$ (for Sa and Sd spirals) of the total circular speed at that radii. The initial rotation curves of these models are given in DM07, Fig. 2.\\

Hereafter, we will adopt the following nomenclature for the different  morphological types: gE0 for giant-like ellipticals, gSa for giant-like Sa spirals, gSb for giant-like Sbc spirals and gSd for giant-like Sd spirals\footnote{We use the term ``giant'' for these systems, to indicate that they are not dwarf galaxies. Their masses are indeed comparable to that of the Milky Way. }. For giant-like Sbc spirals, we will also perform  some simulations adopting a higher gas mass fraction than that  typical of galaxies in the Local Universe. We will use the  nomenclature gSb+ to distinguish these galaxies with a high gas mass  fraction from the "local" Sbc ones.
The complete list of all the  parameters is given in Table \ref{galpar}. We refer the reader to  DM07, for a representation of our galaxy sequence.\\

Since we investigate interactions between giant-like galaxies, the  mass ratios of the interacting systems is always of the order of unity. Intending to run hundreds of simulations, each galaxy is made up  of 120000 particles, distributed among gas, stars and dark matter,  depending on the morphological type (see Table \ref{numbers}). To initialize particle velocities, we adopted the method described in  \citet{hern93}.

\paragraph{Orbital parameters\\}\label{orbital1}

   \begin{table}
      \caption[]{Tree-SPH simulations: Galaxies orbital parameters}
         \label{orbpos}
     \centering
         \begin{tabular}{cccccc}
            \hline\hline
        id  & $r_{ini}$ & ${r_{p}}^{\mathrm{a}}\ $& ${v_{p}}^{\mathrm{a}} \ $ &${E}^{\mathrm{a,b}}\ $ & spin$^{\mathrm{c}}$\\
         & $ \mathrm{[kpc]}$ & $ \mathrm{[kpc]}$& $ \mathrm{[10^2kms^ {-1}]}$ & $\mathrm{[10^4km^2s^{-2}]}$ &\\
            \hline
        01dir &  100. & 8.0 &  7.07 &0.0 & up\\
        01ret &  100.&  8.0& 7.07& 0.0 & down\\
        02dir &  100.&   8.0&  7.42 & 2.5& up\\
        02ret &  100.&  8.0& 7.42& 2.5 & down\\
        03dir &100. &  8.0 & 7.74& 5.0& up\\
        03ret &  100. &   8.0 & 7.74&5.0 & down\\
        04dir &100. &   8.0 & 8.94& 15.0 & up\\
        04ret &  100. &   8.0 & 8.94& 15.0 & down\\
        05dir& 100. & 16.0&  5.00& 0.0 & up\\
        05ret & 100.&16.0&  5.00& 0.0 & down\\
        06dir & 100.  & 16.0 &  5.48&2.5 & up\\
        06ret & 100. & 16.0 &  5.48 & 2.5 & down\\
        07dir &100.& 16.0&  5.92& 5.0 & up\\
        07ret & 100.& 16.0&  5.92& 5.0 & down\\
        08dir& 100.& 16.0& 7.42& 15.0 & up\\
        08ret&100.& 16.0& 7.42& 15.0 & down\\
        09dir& 100.& 24.0&  4.08 &0.0 & up\\
        09ret&100.& 24.0&  4.08 &0.0 & down\\
        10dir& 100.& 24.0&  4.65 &2.5 & up\\
        10ret&100.&24.0&  4.65 &2.5 & down\\
        11dir& 100.& 24.0&  5.16 &5.0 & up\\
        11ret&100.& 24.0&  5.16 &5.0 & down\\
        12dir& 100.& 24.0&  6.83 &15.0 & up\\
        12ret&100.& 24.0&  6.83 &15.0 & down\\
            \hline
         \end{tabular}

\begin{list}{}{}
\item[$^{\mathrm{a}}$] For two equal point masses with $m=2.3 \times10^{11}M_{\odot}$.
\item[$^{\mathrm{b}}$] It is the total energy of the relative motion,  i.e.\\ $E={v}^2/2-G(m_1+m_2)/r$.
\item[$^{\mathrm{c}}$] Orbital spin, if parallel (up) or antiparallel  (down) to the galaxies spin.
\end{list}
   \end{table}

In DM07, seeking to exploit a vast range of orbital  parameters,  we performed 24 different simulations for each couple of  interacting galaxies, varying the orbital initial  conditions, in order to have (for the ideal Keplerian orbit of two  equal point masses of mass $m=2.3\times10^{11}M_{\odot}$) the first  pericenter separation $r_{per}=$ 8, 16, and 24 kpc. For each of these  separations, we varied the relative velocities at pericenter, in  order to have one parabolic and three hyperbolic orbits of different  energy. Finally, for each of the selected orbits, we changed the sign  of the orbital angular momentum in order to study both direct and  retrograde encounters. Combining each orbital configuration with all possible morphologies  for the interacting pair of galaxies, we obtained a total sample of 216 interactions, including only coplanar pairs.

In this work, we have extended the previous sample, taking into  account also different  disk inclinations for the interacting  galaxies. In particular, for each interacting pair in the Tree-SPH sample, we have kept the disk (when present) of one of the galaxies in the orbital plane ($i_1=0^\circ$), and varied the inclination $i_2$ of the companion disk, considering : $i_2=45^\circ, i_2=75^\circ$ and $i_2=90^\circ$,  for a total of 648 new simulations (see Fig.\ref{sketch} for a sketch of the initial orbital geometry and Table \ref{angles} for the coordinates of galaxy spins).

In Table \ref{orbpos}, the initial distance $r_{ini}$ and the  pericenter distance $r_{p}$ between the galaxies center-of-mass are  listed, together with their relative velocity $v_{p}$ at pericenter  and the orbital energy $E$,  for all the simulated encounters\footnote {The values refer to the ideal Keplerian orbit of  two equal point  masses of mass $m=2.3\times10^{11}M_{\odot}$.}.

We have also run 24 additional  simulations of two interacting coplanar ($i_1=i_2=0^\circ$) gSb+ galaxies, for all the possible  orbital parameters given in Table \ref{orbpos}.

In the following, we will refer sometimes to specific encounters, by adopting the morphological type of the two  galaxies in the interaction (gE0, gSa, gSb or gSd), + the encounter  identification string (see first column in Table \ref{orbpos}), +   disk inclination $i_2$ of the second galaxy ($i_1$ is always equal to zero). For  example, the nomenclature gSagSb04ret45 corresponds to an interaction  between a giant Sa and a giant Sb spiral; the Sb disk is inclined of $i_2=45$ degrees with respect to the orbital plane\footnote{For Tree-SPH simulations, $i_1$ being always null, the relative inclination between the two galaxy disks is equal to $i_2$.} and the initial orbital parameters of the encounter are those  corresponding to id=04ret in Table \ref{orbpos}.

\subsubsection{PM-SP simulations}\label{pmspcond}

\paragraph{Galaxy models\\}

\begin{figure}
  \includegraphics[width=8cm]{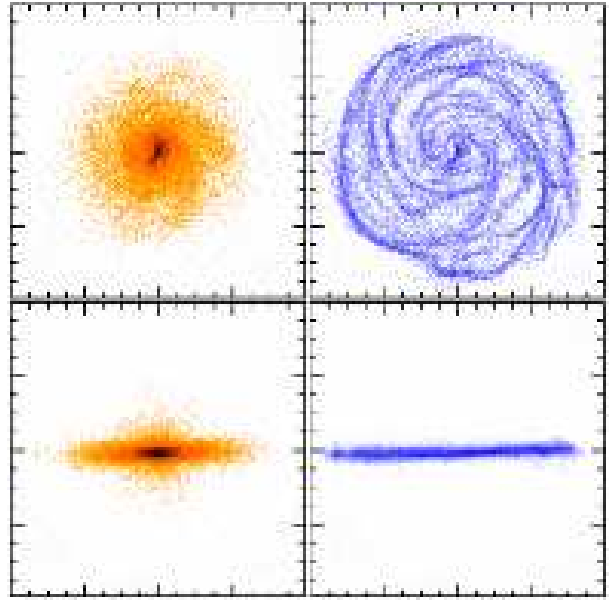} 
\caption{Face-on and edge-on views (left : stars, right : gas) of the ''evolved'' initial conditions in the PM-SP simulations, here for the gSb model galaxy. Each box is 30~kpc~x~30~kpc in size. \label{sbpmspini}}
 \end{figure}

The galaxy mergers in DM07 and in the larger Tree-SPH sample described above do  not show a drastic evolution of the star formation efficiency with  the internal properties of the merging galaxies. In the sample of PM-SP simulations, we model only gSb0-gSb encounters, which are basically  representative on average of any galactic encounter.\\ The gSb-PM model  galaxy is roughly similar to the gSb model from the Tree-SPH sample.  The stellar disk has a Toomre radial profile, the bulge and dark  matter halo a Plummer profile, with mass and scale-lengths indicated  in Table~\ref{galpar}. A  radial scale-length for gas larger than that of Tree-SPH simulations is used. 

The gSb0-PM model has the same parameters than the gSb-PM galaxy, except for the fact that the disk does not contain gas. In other words, all the PM-SP runs concern the interaction of two spiral galaxies, with a bulge-to-disk ratio equal to 0.2, only one of the two disks containing gas.

The galaxy interactions in the Tree-SPH simulations are started with axisymmetric disks (see, for example, Fig.7 in DM07). This is common practice in such numerical studies, however, the rapid transition from  these initial conditions to a more realistic (frequently barred)  spiral distribution occurs during the early phases of the interaction/merger. To make sure that the future conclusions are really related  to the merger-driven evolution and not to this artificial evolution  of the initial conditions, a different method was used in the PM-SP  simulations. The simulated galaxies have been evolved during 1~Gyr in  isolated conditions, before the  simulations of interaction are started.  In this way, the interacting galaxies already have a realistic and  slowly evolving barred spiral structure in the distribution of their  stellar and gaseous components: this helps to make sure that the SFR  evolution relates to merger-driven processes and not to the rapid  formation of the initial spiral structure. 
 This initial 1~Gyr evolution is made over a short period  compared to secular evolution timescales, so that the bulge mass or  disk scale-length do not change dramatically compared to their  initial values. However, star formation is turned off during this  initial phase, so that the gas fraction we have indicated (15~\%)  really corresponds to the gas fraction in the galaxies when the  interaction starts, and any spurious consumption of gas during the  initial formation of the bar and spiral arms is avoided\footnote{In the case of Tree-SPH runs, in turn, simulations are started with axisymmetric disks and star formation is included since the beginning. Thus, in the merging phase, galaxies have yet consumed  a fraction of their gas content (typically 50~\%, even if with large variations, as shown in Fig. 20 of DM07).  Nevertheless, as discussed in that paper, this does not lead to systematic effects on the relative star formation rate of the pairs.}. We show these ''evolved'' initial conditions of the PM-SP models on Fig.~\ref{sbpmspini}.

  \begin{figure*}
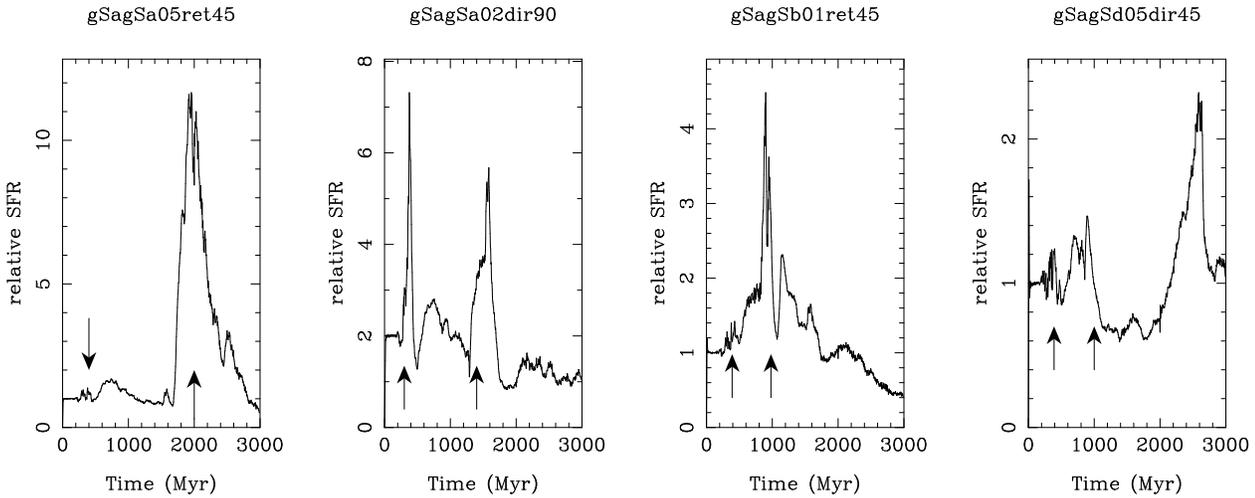

 \begin{minipage}[b]{4.2cm}
   \centering
  \includegraphics[width=6.5cm,angle=270]{sfr3gSagSa05ret45.ps}
 \end{minipage}
\begin{minipage}[b]{4.2cm}
   \centering
  \includegraphics[width=6.5cm,angle=270]{sfr3gSagSa02dir90.ps}
 \end{minipage}
\begin{minipage}[b]{4.2cm}
   \centering
  \includegraphics[width=6.5cm,angle=270]{sfr3gSagSb01ret45.ps}
 \end{minipage}
\begin{minipage}[b]{4.2cm}
   \centering
  \includegraphics[width=6.5cm,angle=270]{sfr3gSagSd05dir45.ps}
 \end{minipage}
\caption{Star formation rate, versus time, for some galaxy mergers. The SFR is normalized to that of the corresponding isolated galaxies. In each panel, the black arrows indicate, respectively, the first pericenter passage between the two galaxies and the merger epoch. \label{sfr}}
\end{figure*}

\paragraph{Orbital parameters\\}

The analysis of the Tree-SPH sample (below) shows that the inclination does not have a major impact on the statistics of  starburst duration and intensity.
 We thus choose not to vary this parameter in the PM-SP  sample, in order to reduce the number of simulations and leave the  possibility to vary the star formation scheme instead. The  inclinations are then fixed to $i_1=33^\circ$ and $i_2=0^\circ$ , because this is about the average value for an isotropic distribution of mergers \citep{bourn07b}. The other orbital parameters are varied as  follows:
\begin{itemize}
\item the encounter velocity $V$ to 50, 100, 150 and 200~km~s$^{-1}$
\item the impact parameter $b$ to 37.5, 50, 62.5 and 75~kpc
\item the orientation to prograde and retrograde.
\end{itemize}
Each combination of these parameters that eventually leads to a merger  of the galaxy pair has been simulated and is included in the  following statistical analysis.

\subsubsection{Core versus cuspy dark matter profiles}\label{nfw}
In both Tree-SPH and PM-SP sets of simulations, the dark matter profiles have been modeled with a core density distribution, which seems to be more in accordance with the rotation curves of local spirals and dwarf galaxies rather than the cuspy halos predicted by Cold Dark Matter (CDM) numerical simulations.\\ Indeed, while cosmological numerical simulations of CDM predict cuspy profiles with a density distribution showing a $\rho \sim r^{-\alpha}$ behavior, with $\alpha \ge 1$, \citep{cole96, navarro96, navarro97, avila98, diemand05}, numerous studies of the rotation curves of local galaxies have shown the observational data are more consistent with the presence of a dark halo having a nearly flat density core ($\alpha \sim 0$). This has been shown to be valid either for dwarf galaxies \citep{flores94, marchesini02, gentile05}, or for Low Surface Brightness (LSB) galaxies \citep{deblok01, mc01, marchesini02, kuzio06, kuzio08}. Some authors have driven the attention to the possibility that systematic effects in the data (as beam smearing, non-circular motions, slit inclinations) could mask the presence of a cusp \citep{van00, van01, swaters03}, suggesting a possible bimodality in the dark halo profiles, with constant density cores restricted to low mass systems ($V_{max} < 70 km/s$) and cuspy profiles for high mass ones \citep{van00}. However, according to \citet{deblok03} none of these systematic effects can reconcile the observational data with cuspy CDM halos. In any case,  as shown for the LSB galaxy UGC4325 by \citet{chemin07}, an accurate modelisation of gas motions seems to be necessary in order to disentangle the dark halo shape of these systems. \\
Moreover, recent results from a new HI mapping of M31  \citep{chemin08} also seem to rule out the presence of a Navarro-Frenk-White cusp in the dark halo profile of this galaxy. \\

The adoption of a cuspy dark halo profile and the induced star formation in merging systems will be the subject of a future paper (Combes et al, in preparation); in any case, some comparisons of bars and gas evolution in isolated galaxies with cuspy Navarro-Frenk-White dark halo profiles or core Plummer ones are discussed in \citet{combes07}. These simulations show that the dark matter profile plays an important role in the development of bar instabilities, and that the presence of a cuspy profile tends to create axisymmetric mass concentrations, which dilutes the gravity torques of the bar and the subsequent gas flows. Similarly, DM cusps may reduce the merger-driven inflows (the  
driving mechanism, gravity torques, being the same as for bar-driven  
flows), which in turn may somewhat decrease the resulting star- 
formation rate. Our choice of core haloes is then in better agreement  
with observations of local spirals, and is also a conservative  hypothesis regarding the intensity of merger-induced starbursts. 

\section{Results from the Tree-SPH simulations}\label{treeresults}

\subsection{Some SFR evolutions}\label{treesfr}

\begin{figure}
  \centering
  \includegraphics[width=9cm,angle=0]{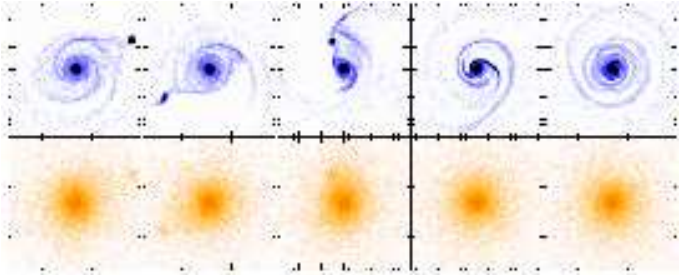}
\caption{Upper panels: gas+new stars maps for the remnant of the merger gSagSd05dir45. Lower panels: old stars maps. From left to right, maps are shown at t=1500 Myr, t=1550 Myr, t=1600 Myr, t=1650 Myr and t=1700 Myr after the coalescence of the gSa and gSb galaxies. Note the presence of a dwarf galaxy, formed during the galaxy encounter, which is falling into the central region of the remnant. Each box size is 12 kpc in length.\label{maps}}
\end{figure}

In this section, we present some SFR evolution during galaxy interactions. In order to distinguish the contribution of tidal effects from secular evolution in determining the star formation history of the two galaxies, all the star formation rates  presented in this and in the next sections are always computed relative to that of the two corresponding isolated galaxies. For the SFRs of the isolated galaxies, they have been discussed and shown  in Sect.4.2 and Fig. 6 of DM07. In the case of Tree-SPH models, galaxy simulations are started with axisymmetric disks, so that the evolution of the isolated systems shows a transient initial burst which is due to compression of the gas into density waves. This transient burst, in turn, is not present in  PM-SP simulations: in this case simulated galaxies have been run in isolated conditions for about 1 Gyr, before the interactions are started.   \\
Fig.\ref{sfr} shows some SFR evolutions with time during major mergers involving galaxies of different morphologies and different orbital parameters. In general, there is a tendency to an enhancement in the star formation rate during an interaction,
 \emph{but}, as it can be seen, a variety of star formation histories are found. In some cases, as in the encounter gSagSa05ret45 (first panel of Fig.~\ref{sfr}), the enhancement of the star formation rate is practically null at the first pericenter passage, while the SFR in the merging phase peaks at more than 10 times that of the corresponding isolated case. The encounter of two gSa galaxies, one having a disk inclined of $90^\circ$ with respect to the orbital plane (second panel of Fig.~\ref{sfr}), leads to two peaks of star formation: the first, at the pericenter passage, being more intense than the one which takes place in the coalescence phase of the two galaxies. The interaction of a gSa and a gSb galaxies, on retrograde orbits, with a relative disk inclination of $45^\circ$ (displayed in Fig.~\ref{sfr}, third panel) causes, in turn, a SFR enhancement which starts at the first pericenter passage and lasts until the final merging phase. Even more striking is the  case shown in the last panel of Fig.~\ref{sfr}, concerning  a giant Sa and a giant Sd moving on a direct orbit (gSagSd05dir45): in this case, the maximum in the star formation evolution takes place neither at the pericenter passage, nor in the merging phase, but about 1.6 Gyr after the coalescence of the two systems. At that epoch, a tidal dwarf galaxy \citep[see][]{duc04}, formed during the first phases of the interaction, falls into the central regions of the remnant, stimulating another episode of star formation (see Fig.\ref{maps} for some maps of this remnant, during the accretion of the dwarf satellite).

\subsection{Starburst frequency}\label{frequency}

   \begin{figure*}
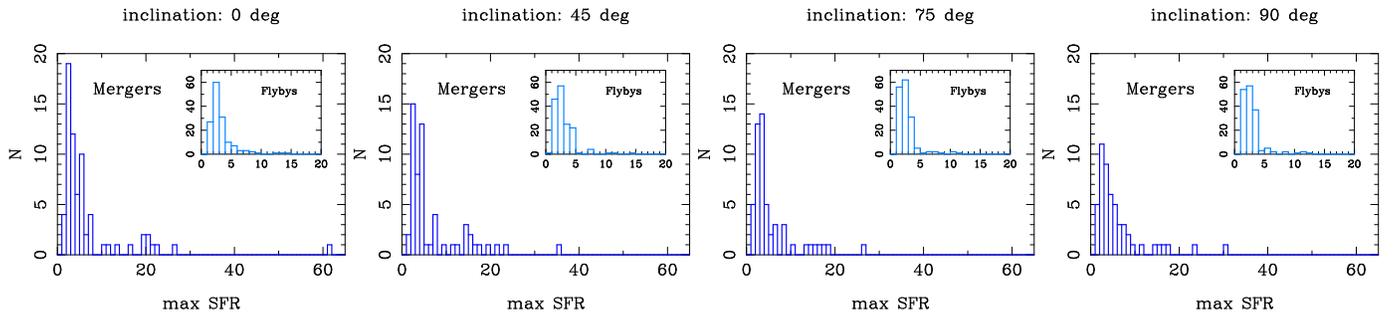

 \begin{minipage}[b]{4.5cm}
   \centering
  \includegraphics[width=4cm,angle=270]{pistogtuttenew00.ps}
 \end{minipage}
 \begin{minipage}[b]{4.5cm}
   \centering
  \includegraphics[width=4cm,angle=270]{pistogtuttenew45.ps}
 \end{minipage}
\begin{minipage}[b]{4.5cm}
   \centering
  \includegraphics[width=4cm,angle=270]{pistogtuttenew75.ps}
 \end{minipage}
\begin{minipage}[b]{4.5cm}
   \centering
  \includegraphics[width=4cm,angle=270]{pistogtuttenew90.ps}
 \end{minipage}
\caption{Histograms of the maximum SFR (relative to the isolated case) for mergers. Flybys are shown for comparison in the small window inserted in the figure. From left to right: histograms relative to encounters with $i_2=0^\circ$, $i_2=45^\circ$, $i_2=75^\circ$ and $i_2=90^\circ$, respectively.  For each inclination, the total number of mergers and flybys is shown in Table \ref{nmernfly}.  More details on the statistical distribution of the maximum SFR are given in  Table~\ref{sfrstat}, Appendix \ref{app2}. \label{istosfr}}
\end{figure*}

   \begin{figure*}
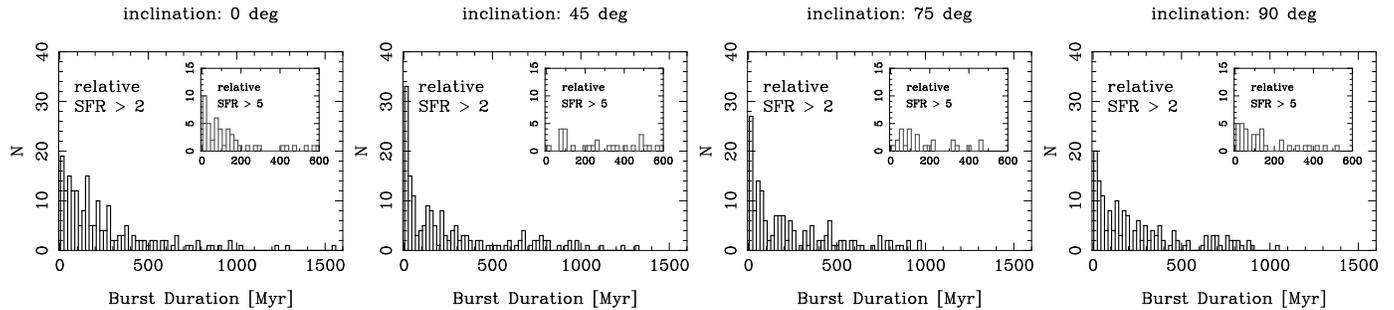

 \begin{minipage}[b]{4.5cm}
   \centering
  \includegraphics[width=4cm,angle=270]{pistotimegtuttenew00.ps}
 \end{minipage}
 \begin{minipage}[b]{4.5cm}
   \centering
  \includegraphics[width=4cm,angle=270]{pistotimegtuttenew45.ps}
 \end{minipage}
\begin{minipage}[b]{4.5cm}
   \centering
  \includegraphics[width=4cm,angle=270]{pistotimegtuttenew75.ps}
 \end{minipage}
\begin{minipage}[b]{4.5cm}
   \centering
  \includegraphics[width=4cm,angle=270]{pistotimegtuttenew90.ps}
 \end{minipage}
\caption{Histograms of duration of enhanced SFR for the whole sample of interacting galaxies (mergers and flybys). Two thresholds are shown: relative SFR $>$ 2 and relative SFR $>$ 5 (the latter is shown in the small window inserted in the figure).  From left to right: histograms relative to encounters with $i_2=0^\circ$, $i_2=45^\circ$, $i_2=75^\circ$ and $i_2=90^\circ$, respectively. More details on the statistical distribution of the duration of enhanced SFR are given in  Table~\ref{timestat}, Appendix \ref{app2}. \label{istotime}}
\end{figure*}

  \begin{figure*}
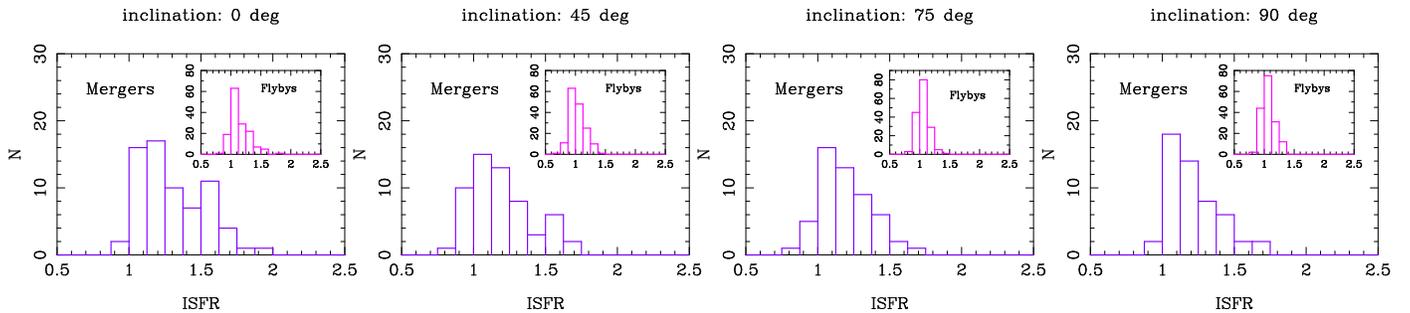

 \begin{minipage}[b]{4.5cm}
   \centering
  \includegraphics[width=4cm,angle=270]{pistoisfrgtuttenew00.ps}
 \end{minipage}
 \begin{minipage}[b]{4.5cm}
   \centering
  \includegraphics[width=4cm,angle=270]{pistoisfrgtuttenew45.ps}
 \end{minipage}
\begin{minipage}[b]{4.5cm}
   \centering
  \includegraphics[width=4cm,angle=270]{pistoisfrgtuttenew75.ps}
 \end{minipage}
\begin{minipage}[b]{4.5cm}
   \centering
  \includegraphics[width=4cm,angle=270]{pistoisfrgtuttenew90.ps}
 \end{minipage}
\caption{Histogram of the relative ISFR for mergers. Flybys are shown for comparison in the small window inserted in the figure. From left to right:  histograms relative to encounters with $i_2=0^\circ$, $i_2=45^\circ$, $i_2=75^\circ$ and $i_2=90^\circ$, respectively.  For each inclination, the total number of mergers and flybys is shown in Table \ref{nmernfly}. More details on the statistical distribution of the ISFR are given in  Table~\ref{isfrstat}, Appendix \ref{app2}. \label{istoisfr}}
\end{figure*}

In the previous section, we have shown that galaxy encounters can lead to a variety of star formation evolutions: some systems show SFR typical of starbust galaxies, while others show only a weak enhancement in the star formation rate during the interaction. In this and in the following sections, we want to deepen the discussion on this point, because of its potential impact on observational and cosmological studies. \\
In particular, in this section, we present some histograms of the maximum star formation rate for mergers and flybys\footnote{By mergers, we mean an encounter which leads to the coalescence of the two galaxies in the simulated time interval (i.e. 3 Gyr), while a flyby is an encounter which does not lead to the coalescence of the two systems during the same time interval.}. As done previously, the SFR is relative to that of the corresponding galaxies evolving isolated. Fig.\ref{istosfr} shows these histograms for three different groups of encounters: those having inclination of $i_2=45^\circ$, $i_2=75^\circ$ and $i_2=90^\circ$ between the disk of one of the two galaxies and the orbital plane. For comparison, coplanar encounters are also shown (the results for coplanar encounters are extensively discussed in DM07). 

\begin{figure*}
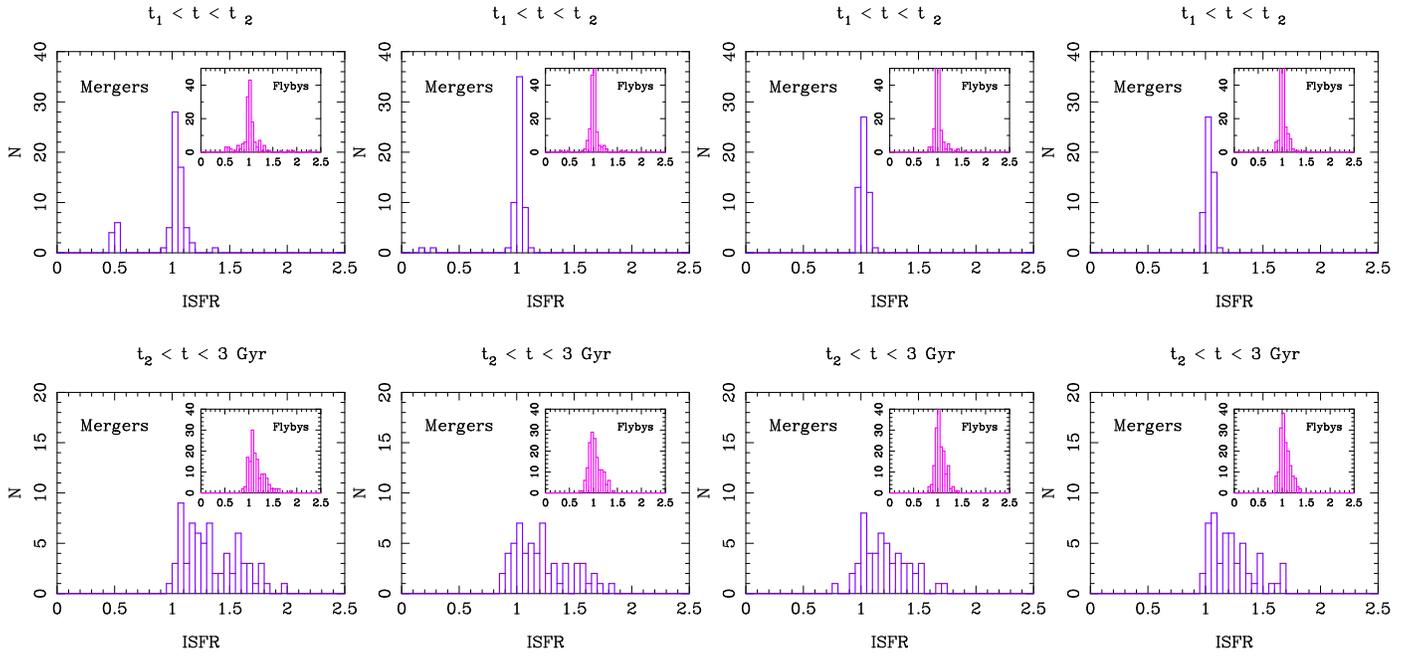

 \begin{minipage}[b]{4.5cm}
   \centering
  \includegraphics[width=4cm,angle=270]{pistoisfr2gtuttenew00.ps}
 \end{minipage}
 \begin{minipage}[b]{4.5cm}
   \centering
  \includegraphics[width=4cm,angle=270]{pistoisfr2gtuttenew45.ps}
 \end{minipage}
\begin{minipage}[b]{4.5cm}
   \centering
  \includegraphics[width=4cm,angle=270]{pistoisfr2gtuttenew75.ps}
 \end{minipage}
\begin{minipage}[b]{4.5cm}
   \centering
  \includegraphics[width=4cm,angle=270]{pistoisfr2gtuttenew90.ps}
 \end{minipage}

\vspace{5mm}
 \begin{minipage}[b]{4.5cm}
   \centering
  \includegraphics[width=4cm,angle=270]{pistoisfr3gtuttenew00.ps}
 \end{minipage}
 \begin{minipage}[b]{4.5cm}
   \centering
  \includegraphics[width=4cm,angle=270]{pistoisfr3gtuttenew45.ps}
 \end{minipage}
\begin{minipage}[b]{4.5cm}
   \centering
  \includegraphics[width=4cm,angle=270]{pistoisfr3gtuttenew75.ps}
 \end{minipage}
\begin{minipage}[b]{4.5cm}
   \centering
  \includegraphics[width=4cm,angle=270]{pistoisfr3gtuttenew90.ps}
 \end{minipage}
\caption{Histogram of the relative ISFR for mergers. Flybys are shown for comparison in the small window inserted in the figure. From left to right:  histograms relative to encounters with $i_2=0^\circ$, $i_2=45^\circ$, $i_2=75^\circ$ and $i_2=90^\circ$, respectively. The top row refers to the ISFR between   $t_1=t_{per} - 50$ Myr and   $t_2=t_{per} + 300$ Myr, being $t_{per}$ the time of the first pericenter passage. The bottom row shows the ISFR in a time $t$ between  $t_2=t_{per} + 300$ Myr and $3$ Gyr. For each inclination, the total number of mergers and flybys is shown in Table \ref{nmernfly}).\label{istoisfrint}}
\end{figure*}

It results that \emph{mergers do not always trigger starbursts}. Indeed the fraction of merging galaxies that produce star formation rates at least ten times higher  than those of isolated galaxies is about $17\%$ of the total merger sample for encounters with inclination $i_2=0^\circ$,  $22\%$  for encounters with inclination $i_2=45^\circ$, while this fraction decreases to $15\%$ and $13\%$ for encounters with inclination of $i_2=75^\circ$ and $i_2=90^\circ$, respectively. Moreover, as the  inclination increases, the  maximum amplitude for the star formation in the merging phase decreases: for coplanar mergers,  the maximum SFR was about 60 times those of the isolated galaxies, while, for polar encounters, it is only 30 times greater than the star formation rate of the isolated systems. Note also that in this figure, as in the case of Figs.\ref{istotime}, \ref{istoisfr} and \ref{istoisfrint}, the number of mergers $N_{mer}$ and flybys $N_{fly}$ changes with the inclination, while the total number of encounters $N_{TOT}=N_{mer}+N_{fly}$ is  equal to 216, for all the inclinations chosen (see Table \ref{nmernfly}).


\begin{table}                                                                                                    
  \caption[]{Number of mergers and flybys in 3 Gyr, the simulated time interval, for the different inclinations $i$.}                                                                                                        \label{nmernfly}                                                                                            
  \centering                                                                                                     
  \begin{tabular}{ccc}                                                                                         
    \hline\hline                                                                                                
    $i$   &  $N_{mer}$ & $N_{fly}$\\
    \hline                                                                                                  
    $0^o$ & 69 & 147\\         
    $45^o$ & 58 & 158\\
    $75^o$ & 53 & 163\\
    $90^o$ & 52 & 164\\
    \hline                                                                                                  
         \end{tabular}                                                                                            
     \end{table}

\subsection{Duration of the star formation enhancement}\label{duration}

Intense starbursts during galaxy mergers are not only less frequent, but also characterized by shorter duration times. In Fig.\ref{istotime}, we present the duration of the star formation enhancement during galaxy mergers and flybys. More precisely, we investigate the duration of the star formation enhancement above a certain threshold, which has been taken equal to two and five times the star formation rate of the isolated galaxies. The histograms in Fig.\ref{istotime} refer only to encounters which sustain a star formation enhancement greater than the chosen thresholds.  
\begin{itemize}
\item Among the 216 encounters\footnote{i.e. including mergers and flybys.} with  inclination $i_2=45^\circ$,  185 sustain a star formation rate which is at least 2 times higher than that of their isolated counterparts (the remaining 31 encounters show no significant enhancement). Among these 185 encounters, only 26 (i.e. only $14\%$) are able to sustain this star formation rate for more than 500 Myr.
\item Among the 216 interactions with  $i_2=45^\circ$, only 46 show a relative star formation enhancement greater than 5 (i.e. about $21\%$ of the total sample). Among these 46 encounters, 20 of them sustain this enhancement for a time greater than 100 Myr.
\end{itemize}

In other words, according to these results, in an interacting galaxy sample, the probability of finding starburst galaxies rather than ``normal'' ones should be small, not only because starbursts are less frequent, as we saw in Sect.\ref{frequency}, but also because the duration of the star formation enhancement shortens as the rate of star formation increases.
Similar trends are found also in the two panels in Fig.\ref{istotime}, which refer to encounters with inclination $i_2=75^\circ$ and $i_2=90^\circ$.

\subsection{Integrated star formation rate}\label{treeisfr}

The results presented in the previous sections suggest that the relation between galaxy interactions and star formation is quite complex. Some systems show in the merging phase a star formation rate which can be ten  times higher than that of isolated galaxies, while many systems show only a weak enhancement. We have also seen that the duration of the enhanced phase of star formation crucially depends on the level of SFR sustained: \emph{the higher the star formation rate, the lower its duration}. The next step is to  quantify the total gas mass converted into stars during an interaction. As in DM07, we quantify the integrated star formation rate during a time interval $T=(t_1,t_2)$ as
\begin{eqnarray}
ISFR=\int_{t=t_1}^{t=t_2}{SFR(t)dt}/\int_{t=t_1}^{t=t_2}{SFR_{iso}(t)dt}
\end{eqnarray}
$SFR(t)$ being the star formation rate of the interacting pair at time $t$ and $SFR_{iso}$ that of the two corresponding galaxies evolving isolated. In this way, it is possible to distinguish secular evolution from the effects due to the tidal encounter. \\
When integrating over the whole duration of the simulations (i.e. over 3 Gyr), 
independently of the  disk inclination $i_2$, we find that mergers do not always convert    high gas mass quantities into new stars (see Fig.\ref{istoisfr}). While coplanar encounters can be efficient enough to produce twice as many stars as isolated galaxies, varying the disk inclination decreases the maximum amount of gas mass transformed into stars: in all  cases, merging galaxies can produce, in the most favorable cases, 1.7 times more stars than their isolated counterparts. But the bulk of merging galaxies shows only a modest enhancement in the ISFR: for $i_2=45^\circ$, for example, $74\%$ of the total merger sample show an $ISFR \le 1.3$ times  the integrated star formation of isolated galaxies. It is also interesting to note that, as the disk inclination increases, moving from $i_2=45^\circ$ to $i_2=90^\circ$, the number of mergers which shows a high value of the ISFR decreases. For example,  mergers with an $ISFR > 1.5$ times the $ISFR$ of the isolated galaxies constitute $14\%$ of the total merger sample with $i_2=45^\circ$, while they are about $7\%$ for mergers with $i_2=75^\circ$ and $i=90^\circ$.\\

In order to understand when most of the gas is transformed into stars during an interaction, we also analyzed the ISFR for two different time intervals:
\begin{enumerate}
\item the first includes the first phase of the interaction, going from $t_1=t_{per} - 50 \mathrm{Myr}$ to   $t_2=t_{per} + 300 \mathrm{ Myr}$, where $t_{per}$ is the time of the first pericenter passage between the two galaxies;
\item the second time interval includes the subsequent phase of the interaction, going from  $t_1=t_{per} + 300 \mathrm{ Myr}$ to $t_2=3 \mathrm{ Gyr}$, the end of the simulation.
\end{enumerate} 


We want to emphasize that, while for mergers, the second time interval includes the coalescence phase of the two galaxies, for flybys most of the tidal effects  act when the galaxies approach one  another (so in the first time interval).\\
The results of this study are shown in Fig.\ref{istoisfrint}. 
When looking at mergers, not surprisingly, we find that most of the conversion from gas to new stars takes place in the second phase of the interaction, for $t_{per}+300 \mathrm{Myr} < t \le 3$ Gyr. Anyway, it should be noted that most of the galaxies in the late phases of the merging process (i.e. in the second time interval chosen) shows only a moderate enhancement in the ISFR: for inclination $i_2=45^\circ$, for example, $69\%$ of the merger sample show an ISFR which is less than 1.3 times that of the corresponding isolated galaxies, and only $19\%$ increase their ISFR of a factor greater than 1.5 with respect to isolated systems. \\
\\  

\begin{figure}
  \centering
  \includegraphics[width=8.8cm,angle=0]{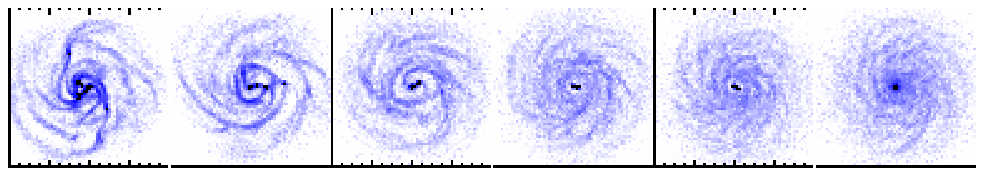}
  \includegraphics[width=8.8cm,angle=0]{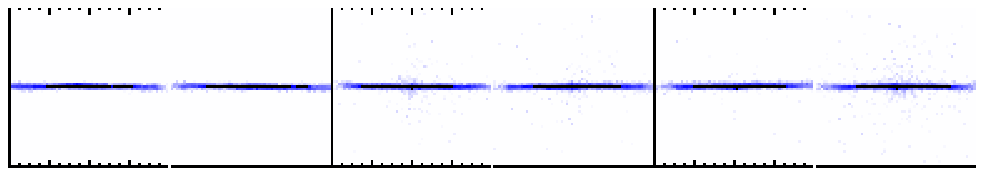}
\caption{Gas maps for the isolated gSb+ galaxy.  From left to right, maps are shown from t=500 Myr to t=3 Gyr,  every 500 Myr. Both xy projection (top panels) and xz projection (bottom panels) are shown. Each box is 40 kpc x 40 kpc in size.\label{isogSb+}}
\end{figure}

Finally, it is interesting to note the response of galaxies to high velocity encounters, which do not lead to the coalescence of the two systems (flybys). Quite surprisingly, these encounters show an enhanced ISFR also in the second time interval, i.e.  well after the pericenter passage, when the systems are at this point well separated. This means that galaxy encounters can stimulate an increase in the star formation rate of a galaxy not only during the phases of close passage, but also well after this phase, when the two systems are far away from each other. This delayed star formation can be caused by the infall of satellite dwarf galaxies formed during the interaction, as shown in Fig.\ref{sfr}, or from instabilities in the galaxy disks, caused by the encounter.

\subsection{Increasing the initial gas fraction}\label{increasing}

\begin{figure}
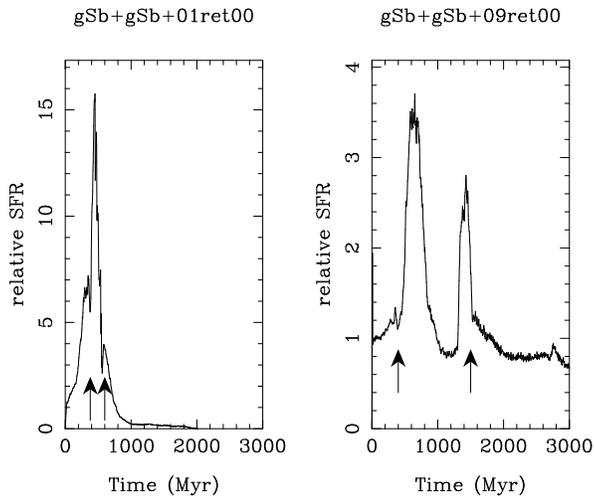

 \begin{minipage}[b]{4.cm}
  \centering
  \includegraphics[width=6.5cm,angle=270]{sfr3gSbgSb01ret00+.ps}
\end{minipage}
\begin{minipage}[b]{4.cm}
\centering
\includegraphics[width=6.5cm,angle=270]{sfr3gSbgSb09ret00+.ps}
\end{minipage}
\caption{Star formation rate, versus time, for some coplanar mergers involving two gSb+ galaxies. The SFR is normalized to that of the corresponding isolated galaxies. In each panel, the black arrows indicate, respectively, 
the first pericenter passage between the two galaxies and the merger epoch.\label{sfr+}}
\end{figure}
\begin{figure}
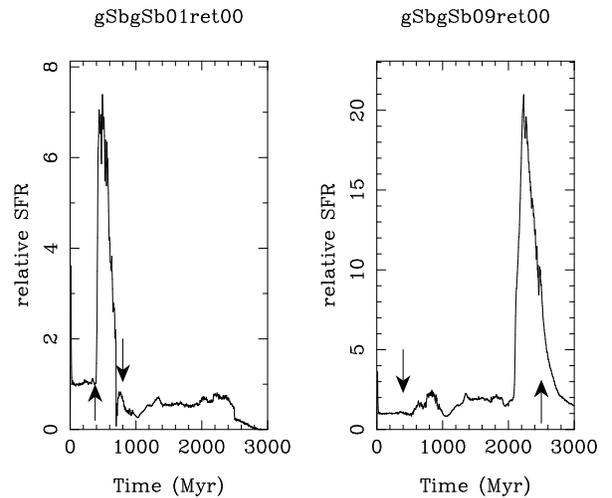

 \begin{minipage}[b]{4.cm}
  \centering
  \includegraphics[width=6.5cm,angle=270]{sfr3gSbgSb01ret00.ps}
\end{minipage}
\begin{minipage}[b]{4.cm}
\centering
\includegraphics[width=6.5cm,angle=270]{sfr3gSbgSb09ret00.ps}
\end{minipage}
\caption{Star formation rate, versus time, for some coplanar mergers involving two gSb galaxies. The SFR is normalized to that of the corresponding isolated galaxies. In each panel, the black arrows indicate, respectively, 
the first pericenter passage between the two galaxies and the merger epoch.  See Fig.\ref{sfr+} for a comparison of these SFRs with those of gSb+ galaxies, having the same orbital parameters.\label{sfrno+}}
\end{figure}

\begin{figure*}
  \centering
  \includegraphics[width=18cm,angle=0]{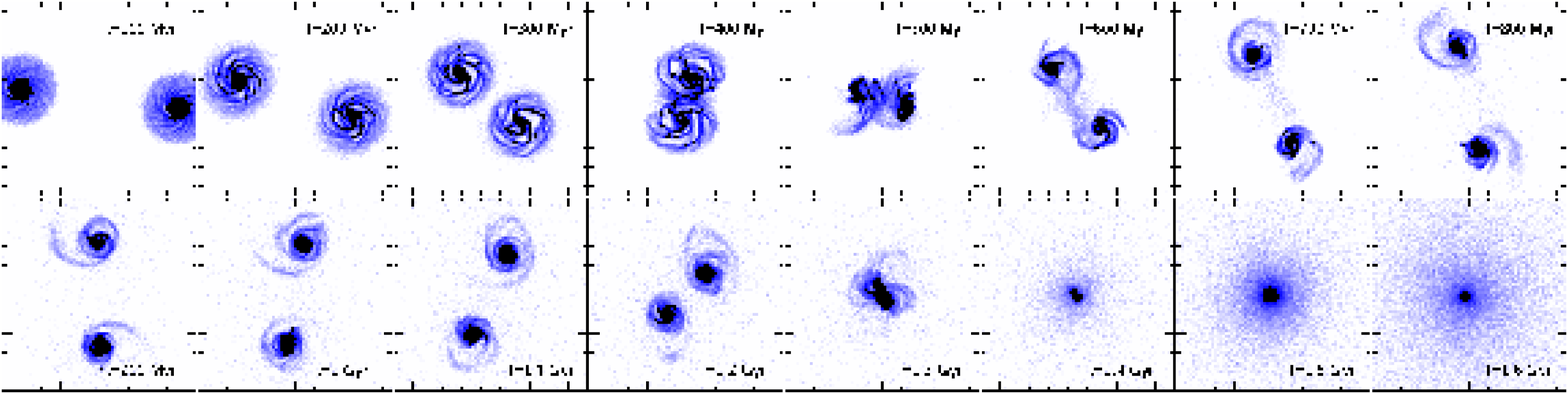}
  \includegraphics[width=18cm,angle=0]{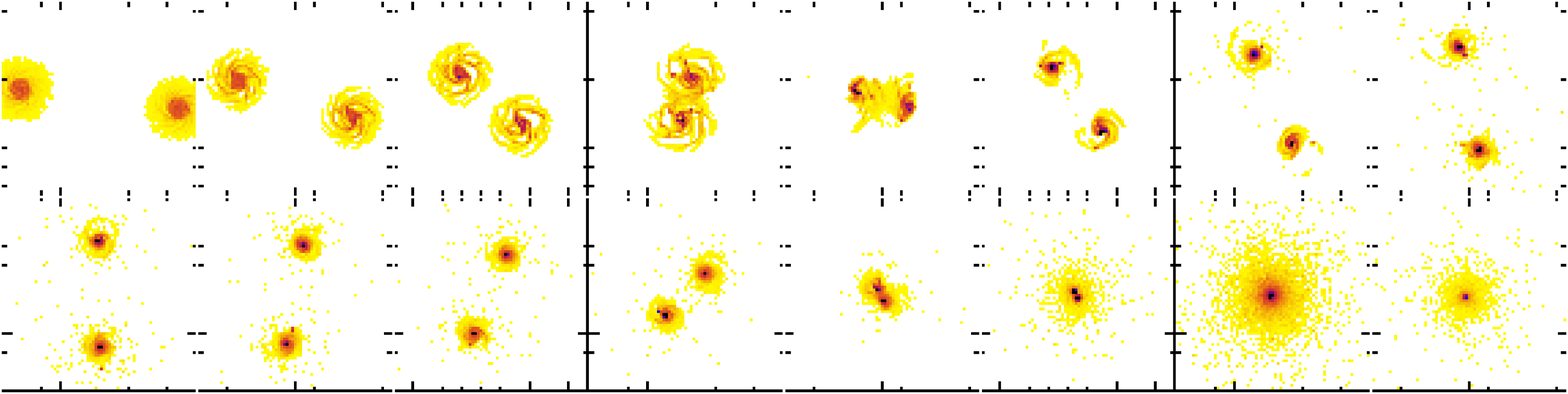}
\caption{Gas maps (top) and star forming regions (bottom) for a retrograde merger involving two gSb+ galaxies (id=gSb+gSb+09ret00). The corresponding SFR is shown in Fig.\ref{sfr+} (right panel). Each panel is 100 kpc x 100 kpc in size.\label{gasgSbgSb09ret00+}}
\end{figure*}

All the results presented in the previous sections concern the interaction of galaxies having a gas mass content  typical of galaxies in the Local Universe. In this section,  we want to discuss how and to what extent our results depend on the total amount of gas available in the galaxy disk. In other words, how does the star formation rate change, when the gas mass fraction in the galaxy increases? To answer this question, we have run a small number of simulations (24) of coplanar interactions between two Sb galaxies, having an initial gas mass equal to one half that of the stellar disk\footnote{Hereafter called gSb+ galaxies, to distinguish them from the gSb galaxies, presented so far.}. \\
The initial conditions chosen for the gSb+ galaxies are such that the initial disks are stable (see Fig.\ref{isogSb+} for some gas maps of the gSb+ galaxy, evolving isolated for 3 Gyr).  For that, we have increased the initial value of the Toomre parameter $Q_{gas}$ (see Table \ref{galpar}). Probably, galaxies at higher redshifts are more unstable, and the instability of their disks can give rise to intense star formation (see for example \citet{elm07}), via fragmentation of the gas disk and clumps formation \citep{bourn07a}. 
We also kept unchanged the extension of the stellar and gaseous disks, even if observations suggest that the sizes of disks decrease significantly with redshift \citep{maoetal98, giallo99, avi01}. These choices are all consistent with the fact that we want to  investigate the effect of increasing the gas fraction in the disk, rather than accurately modeling galaxies at higher redshifts.   \\

Some evolutions of the star formation rate, as a function of time, for gSb+-gSb+ interactions, are given in Fig.\ref{sfr+}. As previously done, the SFRs are normalized to those of the corresponding isolated galaxies (but see Appendix \ref{absolutesfr} for the evolution of some absolute SFRs).\\ The left  panel in this figure presents the star formation evolution during a retrograde merger (id=gSb+gSb+01ret00): in this case, the coalescence of the two galaxies takes place only 200 Myr after   the first pericenter passage. The stimulated star formation rate, after the first encounter, is quite high (about 15 times that of the isolated galaxy), while its amplitude is reduced in the merging phase, when the SFR is only 4 times that of the isolated counterparts. \\The response of the star formation evolution to tidal effects is also shown in the case of the retrograde merger with id=gSb+gSb+09ret00 (right panel in Fig.\ref{sfr+}). In this case, after the first passage (t=400 Myr), the star formation rate increases up to about 3.5 times that of the corresponding isolated galaxies (t=600 Myr), then it declines to preinteraction levels, to rise finally in the merging phase up to about 3 times the corresponding value of the isolated systems.\\

Comparing these SFR evolutions with those of gSb galaxies, having the same initial orbital conditions, quite surprisingly we find that in the merging phase the SFR (relative to the isolated case) is higher for local gSb galaxies, than for gSb+ systems, which have initially a higher gas mass fraction (cf, for example, right panels in Fig.\ref{sfr+} and \ref{sfrno+}). In turn, the amplitude of the first SFR peak, just after the first close passage between the two systems, is higher in the case of gas-rich interactions than for their gas-poor counterparts. These  trends are common to all the simulations performed with the Tree-SPH code. \\

\subsubsection{Disk fragmentation}

\begin{figure}
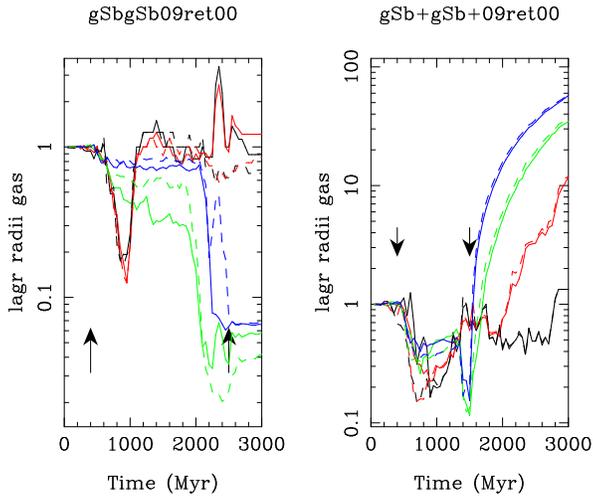

 \begin{minipage}[b]{4.cm}
  \centering
  \includegraphics[width=6.5cm,angle=270]{plagrgasgSbgSb09ret00.ps}
\end{minipage}
\begin{minipage}[b]{4.cm}
\centering
\includegraphics[width=6.5cm,angle=270]{plagrgasgSbgSb09ret00+.ps}
\end{minipage}
\caption{Evolution with time of the gas lagrangian radii containing $10\%$ (black), $25\%$ (red), $50\%$ (green) and $75\%$ (blue) of the gas mass, in units of the corresponding gas radii of the isolated galaxies. Left panel: local gSbgSb09ret00 interaction; right panel: gas-rich gSb+gSb+09ret00 encounter. In both panels, the lagrangian radii of each of the two galaxies involved in the interaction are shown (solid and dashed lines, respectively). \label{lagr}}
\end{figure}


In this section, we want to understand the mechanism responsible of  stimulating higher levels of star formation at the first pericenter passage in gas-rich systems rather than in local ones. The first hypothesis that we have tested is that the high star formation rates found in gas-rich systems  could be due to a more pronounced gas inflow in the central regions. In order to understand if this is the mechanism at work, we evaluated the radii $r_{10}, r_{25}, r_{50}$ and $r_{75}$ containing, respectively, $10\%, 25\%, 50\%$ and $75\%$ of the gas mass present in the local interacting system with id=gSbgSb09ret00 and in the corresponding gas-rich pairs gSb+gSb+09ret00. The evolution with time of these radii is shown in Fig.\ref{lagr}. As usual, in order to separate the effects due to the tidal interaction from secular evolution, the radii are normalized to that of the corresponding isolated systems. In both cases (local and gas-rich systems), it is evident that before the first pericenter passage, all these values are unchanged with respect to the isolated systems (indicating that at these times, the interaction is still not efficient in perturbing the gas dynamics). But then, just after the pericenter passage (t=400 Myr), the disks of the two gSb galaxies and of their gas-rich gSb+ counterparts are quite perturbed: this gives rise to a compression of the inner radii ($r_{10}, r_{25}$) of a factor of ten in both cases (see, respectively, left and right panel in Fig.\ref{lagr}). \emph{But} the gas is compressed by the same amount in both pairs. This means that the inflow could not be responsible, alone, of driving different star formation rates, relative to the isolated systems, at this time: indeed the same amount of compression for local and gas-rich systems should stimulate the same amount of star formation enhancement.

The important difference between the local interaction and that involving gas-rich systems resides in the fact that these latter, being more unstable, are more prone to fragment under the tidal effects of the companion galaxy. In this sense, for these simulations, \emph{we should talk of fragmentation-driven starbursts rather than inflow-driven ones}. In Fig.\ref{gasgSbgSb09ret00+}, for example, it is clear that the first close passage of the two galaxies is accompanied by the formation of many gas clumps in the disks, which are also site of intense star formation. This fragmentation is less evident for the gSb+ galaxies evolved in isolation, thus indicating that this disk instability is largely a result of the encounter. Note also that the formation of gas clumps in the galaxy disks was less evident for local interacting  pairs (see maps in DM07). As we discuss in more details in Appendix \ref{app1}, in the case of some isolated gas-rich galaxies, the more important is the disk fragmentation, the higher is the star formation rate in the galaxy. This indicates that the disk fragmentation can strongly contribute to the great SFR enchancement  found for gas-rich spirals at the pericenter passage.

\subsubsection{Effects of feedback}                                                                                                                           
If, on the one side, gas-rich systems present higher star formation enhancements than their local counterparts at the pericenter passage, on  the other side, in the merging phase, gas-rich galaxies show in general lower star formation rates\footnote{Relative to the corresponding isolated case.} than local interacting pairs (see Figs.\ref{sfr+} and \ref{sfrno+}). \\      What is the physical mechanism that can inhibit a starburst in the merging phase of these gas-rich gSb+ spirals? As we checked, before the coalescence of the two systems, the gas amount is still enough to produce a considerable enhancement of the star formation rate. So, why is this gas reservoir not sufficient to drive important bursts of star formation? To answer this question, first of all we analyzed the evolution of the gas component  during the gSb+gSb+09ret00 merger. Fig.\ref{gasgSbgSb09ret00+} shows some maps of gas and star forming regions  during the interaction. Since the spin of the two galaxies is antiparallel to the orbital angular momentum (retrograde interaction), the morphology of the two disks is only sligthly affected after their first close passage: no long tidal tails are formed,  only a bridge connects the two systems. Note also that at that epoch (t=600 Myr), just after the first close passage, the SFR reaches its maximum value. As it can be seen from the maps, in this phase, the star formation is quite extended along the two disks, from the central regions to spiral arms. As the two galaxies begin to approach one another once again, the gaseous component tends to collapse in the central disk regions (t=1.3 Gyr - 1.4 Gyr), thus giving rise to the SFR enhancement found in Fig.\ref{sfr+}. But just after this strong inflow, the gas expands, reducing drastically the 
star formation rate. \\                                                                                                                                       
\begin{figure}                                                                                                                                                
  \centering                                                                                                                                                  
  \includegraphics[width=5cm,angle=270]{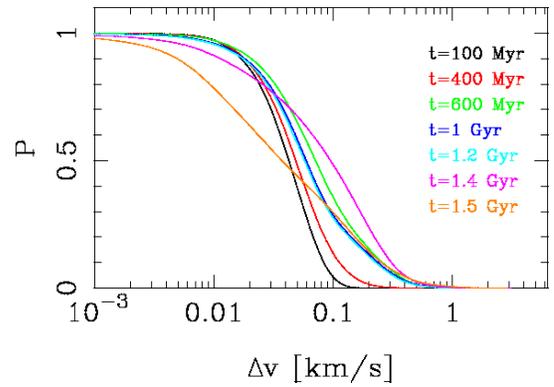}                                                                                                      
\caption{Probability to have a kinetic feedback larger than the value specified on the x-axis for gas particles, during the encounter with id=gSb+gSb+09ret00\
. Different curves correspond to different epochs during the interaction, as explained in the figure.\label{feed}}                                            
 \end{figure}                                                                                                                                                 
                                                                                                                                                              
The response of the gaseous component after the strong inflow is a consequence of the kinetic feedback released by supernovae explosions.  As described in DM
07, energy injection into the ISM is included following the work by \citet{mih94b}, by applying a radial kick to velocities of neighboring gas particles. In 
DM07, performing a set of simulations of isolated galaxies, we found that a rapid growth of the vertical thickness of the gaseous disk can be prevented if  the total amount of kinetic energy received by a gas particle, due to the contribution from all neighbors, corresponds to less than 1 km/s. Fig.\ref{isogSb+} 
clearly shows that this is the case also for gas rich systems: indeed, with such an amount of feedback, the vertical thickness of the gaseous disk remains quite constant during the evolution.\\                                                                                                                             
Evidently, the amount of feedback acquired by a gas particle depends on the local star formation rate: particles that lie in a actively star forming region receive an energy injection greater than that acquired by particles in the disk outskirts.\\                                                                   
Galaxy interactions, perturbing the star formation evolution, also change the amount of kinetic feedback received by the ISM. Fig.\ref{feed} shows the probability of finding gas particles receiving an amount of feedback greater than a certain value, in an interval of values ranging from $10^{-3}$ to $6\rm{kms^{-1}}$, for the gSb+gSb+09ret00 encounter. In this plot, for each particle, the kinetic feedback has been evaluated as the total velocity impulse $\Delta v$ received by a gas particle, due to the contribution from all its neighbors.\\                                             
In the early phases of the interaction (t=100 Myr), when the two systems are sufficiently far away from each other to be considered isolated, 50$\%$ of the gas particles receive a total radial velocity kick from neighbors which is less than 0.03 $\rm{kms^{-1}}$, only $1/20$ having $\Delta v$ greater than 0.1  $\rm{kms^{-1}}$. As the interaction proceeds, and the two galaxies approch the pericenter (t=400 Myr), the amount of kinetic energy received increases. This trend continues until t=600 Myr, when the peak of SFR occurs. But the strongest increase in the kinetic feedback of the system is obtained in the merging phase: at t=1.4 Gyr, indeed, 50$\%$ of the gas particles receive a velocity kick greater than 0.1  $\rm{kms^{-1}}$. Note that such a high feedback for such a high fraction of gas particles is due to the strong inhomogeneity of the star forming regions. In other words, the strong gas inflow produces a highly concentrated star forming region, where gas particles acquire conspicuous radial kicks from neighbors.                                                                    
This rise in the kinetic energy determines the gas expansion found in Fig.\ref{gasgSbgSb09ret00+} and the subsequent decrease in the star formation rate of the merger.                  

\section{Comparison with PM-SP simulations}\label{comparison}

Are the main results found for Tree-SPH simulations still valid for PM-SP ones? Or, in other words, to what extent do the results presented in the previous section depend on the numerical techniques adopted? Let us first recall the various assumptions that differ between the Tree-SPH and PM-SP models, and could a priori cause major differences in the star formation history:
\begin{itemize}
\item $N$-body and gas dynamics schemes are completely different (tree- versus grid-based).
\item The initial conditions differ: axisymmetrical in the Tree-SPH sample, with evolved spirals in the PM-SP models. Dark matter extents are also different.
\item A local Schmidt law was used in both models, but applied in a different way: in the Tree-SPH simulations, the gas density is computed using the adaptative tree structure, implicitely assuming a scale-free Schmidt law, while the gas density is computed on a grid with constant resolution in the PM-SP simulations.
\item Different star formation schemes were used only in the PM-SP models.
\item 
One of the galaxies in the pair is always gas-free in the PM-SP models, while different morphologies (and gas contents) have been adopted for Tree-SPH simulations.
\end{itemize}

Some other differences are not expected to influence much the star formation:
\begin{itemize}
\item The orbital inclination is not varied in the PM-SP models, but according to the Tree-SPH models this parameter has little influence on the statistical properties of star formation (Figs.~\ref{istosfr} to \ref{istoisfr}).
\item Star formation feedback is included only in the Tree-SPH models.
\end{itemize}
\medskip

We now discuss the robustness of the main results obtained from the Tree-SPH models by comparing them with the results of PM-SP simulations. Keeping in mind that these two samples contain interactions with different orbital parameters (impact parameter, relative velocities, disk inclination), to compare them properly we have considered that encounters have different likelihoods, with a probability $p \propto b^2 V$, $b$ being the impact parameter of the encounter and $V$ the relative velocity of the two systems. This choice is motivated since the collision rate is proportional to the relative velocity $V$ and to the cross-section $\pi b^2$. We have also tested other possibilities, including  a dependence on the inclination  $i$ of the galaxy disks (for example, by means of a factor $ \propto sin(i)$), but this did not change significatively the results. We refer the reader to \citet{bourn07b}, \citet{mart07} and  \citet{mihos04} for a further discussion on  statistical weighting of interacting galaxy models.\\ 
Results obtained this way, shown on Figs.~\ref{cum_sfr} and \ref{cum_time}, allow a statistical comparison of the two numerical datasets, as well as a statistical comparison of simulations with observations.

\subsection{Frequency and intensity of merger-induced starbursts}
\begin{figure}
 \begin{minipage}[b]{7.2cm}
  \centering
  \includegraphics[width=6.5cm,angle=270]{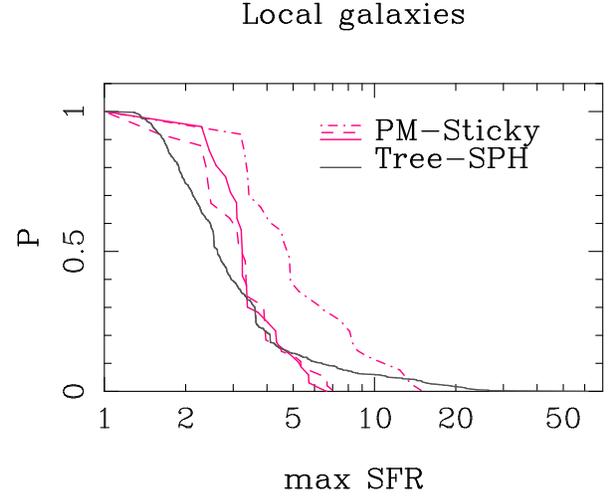} 
   \end{minipage}
\caption{Probability to have a maximum SFR larger than the value specified on the x-axis for major mergers. The SFR is normalized  to that of the corresponding isolated galaxies. The grey curves correspond to Tree-SPH simulations, the red curves to Particle mesh-sticky particles ones. Different lines correspond to simulations employing different recipes for modeling star formation: $\Sigma_{SFR} \propto {\Sigma_{gas}}^{1.5}$ (solid lines), $\Sigma_{SFR} \propto {\Sigma_{gas}}\Omega$ (dashed line), $\Sigma_{SFR} \propto {\Sigma_{gas}}^{1.5}\Omega$ (dot-dashed line). \label{cum_sfr}}
 \end{figure}

\begin{figure}
 \begin{minipage}[b]{7.2cm}
  \centering
  \includegraphics[width=6.5cm,angle=270]{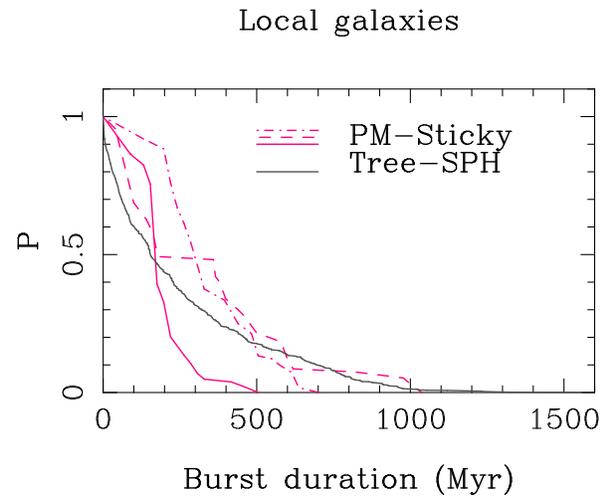} 
   \end{minipage}
\caption{Probability to have a star formation rate larger than two times the star formation of the isolated galaxies for a time  larger than the value specified on the x-axis during major galaxy mergers. The grey curves correspond to Tree-SPH simulations, the red curves to Particle mesh-sticky particles ones. Different lines correspond to simulations employing different recipes for modeling star formation: $\Sigma_{SFR} \propto {\Sigma_{gas}}^{1.5}$ (solid lines), $\Sigma_{SFR} \propto {\Sigma_{gas}}\Omega$ (dashed line), $\Sigma_{SFR} \propto {\Sigma_{gas}}^{1.5}\Omega$ (dot-dashed line). \label{cum_time}}
 \end{figure}

Comparing the statistical distribution of the maximum star formation enhancement\footnote{As usual, relative to the star formation of isolated systems.} of local interacting pairs, it is striking to see how the results are in agreement, even adopting different numerical approaches. Indeed, Tree-SPH simulations and PM-SP ones with identical star formation recipes (Schmidt law, only density-dependent) have very similar distributions (see Fig.\ref{cum_sfr}): in both cases, it results  that \emph{interactions and mergers, in general, produce moderate enhancements in the star formation rate of the pairs, while  strong starbursts are rare}. Indeed, less than 15$\%$ in each set of simulations show a maximum star formation rate greater than 5 times that of isolated systems. The median value of the SFR enhancement is 2.8 in the Tree-SPH dataset, and 3.1 in the PM-SP dataset. 

Moreover, these results are not dependent on the models adopted to compute local star formation rates, as shown by the comparison of the various subset of PM-SP simulations. The $\Sigma_{SFR} \propto {\Sigma_{gas}}\Omega$ model (compatible with observations, see Sect.~\ref{recipe} and Fig.~\ref{sfrobs}) indeed shows a statistical distribution quite similar to the Schmidt law model. Higher probabilities of strong starbursts are found with the $\Sigma_{SFR} \propto {\Sigma_{gas}}^{1.5}\Omega$ model: as discussed in Sect.\ref{recipe} (and Fig.~\ref{sfrobs}), this parametrization is barely compatible with observations and should be considered as an upper limit to the actual starburst efficiency, hence confirming our earlier conclusions. Even in this favorable case, the majority of the encounters (60$\%$) leads to a maximal star formation enhancement smaller than a factor of five.

Starbursts stronger than a factor 5 compared to the isolated galaxies are rare for the most realistic star formation models (Schmidt law and $\Sigma_{SFR} \propto {\Sigma_{gas}}\Omega$ model) in both the Tree-SPH and PM-SP simulations. However, in more detail, one can note a difference for very high efficiencies: some starbursts reach factors 10--20 in the Tree-SPH dataset, while no similar case is found in the PM-SP dataset. An explanation might be that the PM-SP sample is smaller and has randomly missed these most efficient cases, however the sample should have been large enough to obtain a few cases in this range. This difference must then have a physical cause, which could be one of the following:
\begin{itemize}
\item Some peculiar and very favorable conditions are not included in the PM-SP sample. In particular, perfectly coplanar encounters between two gas-rich galaxies can enhance the star formation rate by direct collision between the two gas components when the two disks overlap: this will create high density regions and dissipate the gas angular momentum very rapidly, triggering the gas infall towards the center. Such situations should, however, be rare and correspond only to very specific situations.
\item This could also result from the different application of the Schmidt law: in the Tree-SPH code, the Schmidt law is assumed to be scale-free and the gas density is computed on the tree structure. In the PM-SP code, the gas density is computed at a fixed scale on the Cartesian grid. Very dense small-scale structures will be resolved only in the first case, while their density will be smoothed at the grid scale in the second case. Which model is the most realistic depends on the physical origin of the Schmidt law. If this law comes from the turbulent density distribution in the ISM \citep{elmegreen02} it should be scale-free just like in the Tree-SPH code. If the Schmidt law results from the dynamical timescale of gas clouds $1/\sqrt{G \rho}$ \citep{elmegreen02}, it should not be scale free, and the density should be computed when clouds begin to form on rather large scale, not at smaller scales when they have begun to collapse. In this case, the Tree-SPH may over-estimate the star formation efficiency by resolving density increases at small scales.
\end{itemize}

 \begin{figure}
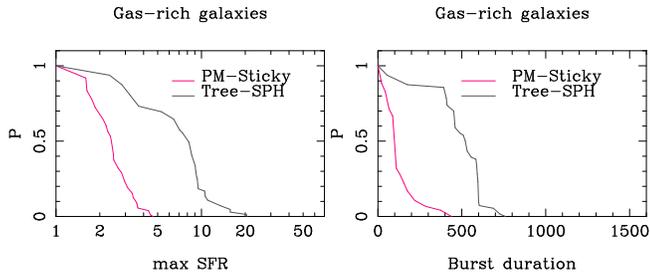

  \begin{minipage}[b]{4.2cm}
  \includegraphics[width=3.5cm,angle=270]{sfr_comp_r2+gas_pesoeqr2v.ps}
   \end{minipage}
 \begin{minipage}[b]{4.2cm}
  \includegraphics[width=3.5cm,angle=270]{time_comp_r2+gas_pesoeqr2v.ps}
   \end{minipage}
\caption{Same as Figs.~\ref{cum_sfr} and \ref{cum_time} for very gas-rich galaxies (gas fraction in the disk: 50\%). The grey curves correspond to Tree-SPH simulations, the red curves to Particle mesh-sticky particles ones. Large star formation rates in Tree-SPH simulations are attributable to instabilities resulting in the collapse of massive star-forming clumps during the interaction (see text). The effect of this internal fragmentation was removed in PM-SP simulations so that star formation is driven only by the merger-induced gravity torques; starbursts are then much more modest. 
 \label{cum_gas}}
\end{figure}

Some differences concerning the most efficient starbursts can thus be caused either by different numerical techniques, or by different physical assumptions on star formation. They nevertheless affect only rare cases or have limited amplitudes. We note in particular that, for the moderate starbursts up to a factor $\simeq 4$ in the relative SFR (that are the most frequent), the PM-SP model finds higher average efficiencies than the Tree-SPH one (Fig.\ref{cum_sfr}). This could be caused by the absence of feedback in the PM-SP model, as is indeed expected that feedback should regulate the star formation activity. Moreover, the differences expected by including feedback or not in the two models might be somewhat compensated for by some other assumptions: in the PM-SP model, not including feedback favours an increase of the star formation activity, but on the other hand evaluating the gas density on the fixed resolution grid smoothes the density peaks and should somewhat regulate star formation compared to the Tree-SPH method.

Overall, the Tree-SPH and grid-sticky particle simulations agree that strong starbursts are not frequent in major galaxy mergers, with a median starburst efficiency of about a factor 3 compared to isolated galaxies. This conclusion is robust versus numerical methods. It is also robust versus morphological types of merging galaxies (from the Tree-SPH sample) and the physical assumptions on the model used to compute local star formation rates (from the three PM-SP subsamples).

\subsection{Duration of merger-induced starbursts}

The duration of the  starbursts triggered by galaxy mergers is found to be generally smaller than 500~Myr in the Tree-SPH models. The PM-SP models have a somewhat different statistical distribution of the starburst duration (see Fig.~\ref{cum_time}) that also depends on the adopted star formation model. However, both codes and all star formation models are in agreement on the main conclusions: no more than 20\% of interaction/merger-induced starbursts have durations longer than 500~Myr, the Schmidt law in PM-SP models giving even no such case. The Schmidt law as well as the $\Sigma_{SFR} \propto {\Sigma_{gas}}\Omega$ have median durations of 200~Myr, and average durations of 200--300~Myr. This is obtained for disk stellar masses of several $10^{10}$~M$_{\sun}$, thus durations should be even shorter for smaller disks or higher redshift disks that have shorter dynamical timescales. Also note that, as discussed in Sect.~\ref{duration}, the burst duration diminishes as the star formation enhancements increase, so that the values found in the left panel of Fig.\ref{cum_time} should be taken as upper limits if they are compared to observations. Let us note that observations are biased towards the strongest starbursts, because of brightness selection effects.

\subsection{The case of gas-rich galaxies}\label{comp_gas}

\begin{figure}
  \includegraphics[width=8cm]{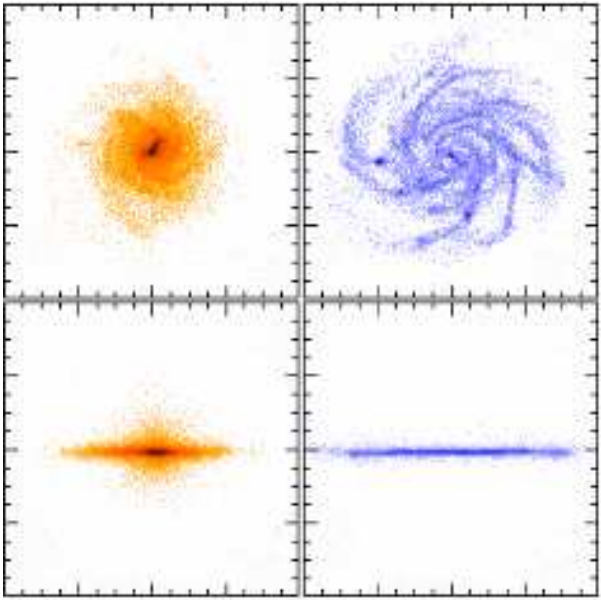} 
\caption{Face-on and edge-on views (left : stars, right : gas) of the initial conditions for the gSb+ gas-rich galaxy in the PM-SP models (each box is 30~kpc~x~30~kpc in size). Note that gas clumps here {\it (i)} already exist in the initial conditions while they form during the merger in the Tree-SPH simulations and {\it (ii)} these clumps are smoother, less compact and less massive than those formed in the Tree-SPH models, owing to the reduced dissipation rate and non-adaptative resolution.\label{gsbpmspini}}
 \end{figure}

\begin{figure*}
  \includegraphics[width=18cm]{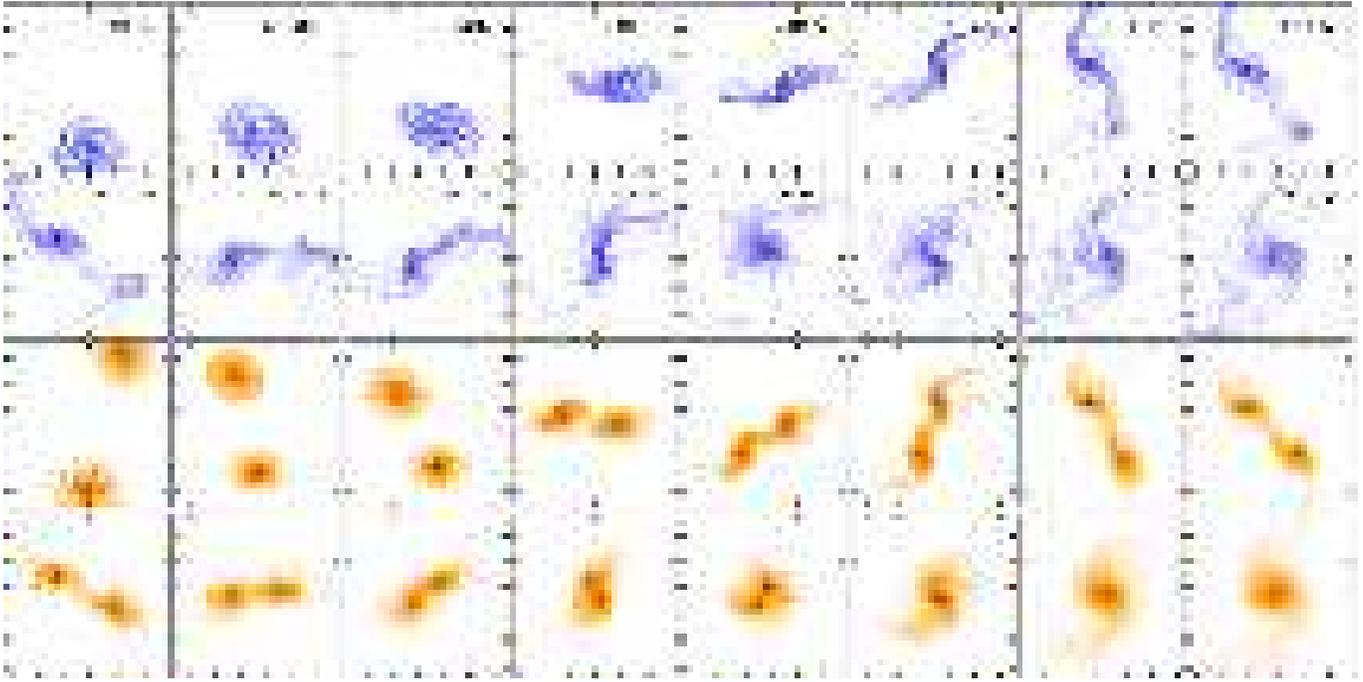} 
\caption{Gas (top) and stars (bottom) maps for a gSb0-gSb+ merger (with an impact parameter of 62.5~kpc and an encounter velocity of 150 km~s$^{-1}$) for the PM-SP model. Each box is 70~kpc~x~70~kpc in size. \label{gsbpmsp}}
 \end{figure*}

As discussed above, the Tree-SPH and PM-SP simulations show a good agreement when modeling local
 galaxy interactions. The gas fraction, varied from 10 to 30\% in the Tree-SPH simulations, does not seem to have a major impact on the relative intensity of merger-induced starbursts: disks that contain more gas have higher SFRs both before and during mergers, and the ratio of enhancement is more or less unchanged (see also DM07).

For particularly gas-rich systems (gSb+ models, assumed to represent gas-rich high-redshift galaxies),  the situation is much more complex, as shown in Fig.\ref{cum_gas}. The PM-SP models show roughly unchanged distribution of the starburst relative efficiency and duration, while the Tree-SPH models show an increase of both the efficiency and duration of starbursts. As we saw in Sect.~\ref{increasing} in the case of Tree-SPH simulations, it is mainly in the first phases of the interactions that gas-rich pairs show a higher star formation rate than galaxies with lower gas fractions, while  in the late merging phase star formation activity is lowered when the gas fraction is increased, due to stronger feedback, or earlier gas consumption effects. We have also seen that, for gas-rich spirals, the star formation enhancement at the pericenter passage is mainly due to disk instabilities and clumps formation, rather than to stronger gas inflows. In other words, for gas-rich systems, it is a ''merger-driven fragmentation' rather than ''merger-driven inflows'' that drives the star formation  in the Tree-SPH simulations with 50\% gas. The star formation enhancement driven by disk fragmentation is also responsible of the longest burst duration found in the right panel of Fig.\ref{cum_time} (a more detailed discussion on this point can be found in Appendix).

The different behaviour of PM-SP simulations is due to the fact that in order to prevent an excessive disk fragmentation in the galaxy evolved isolated, the gas dissipation in these simulations had been decreased (see Section 2.1). This essentially prevents the formation of dense gas clumps, sites of intense star formation. Actually, the gas-rich PM-SP models are shown in Fig.~\ref{gsbpmspini} for the starting conditions and Fig.~\ref{gsbpmsp} for a merger sequence: some clumps of gas are also present, but owing to the reduced dissipation they are less massive and less dense that those seen in the Tree-SPH simulations. Moreover, they are present in the starting conditions and not formed during the merger. There is no additional fragmentation during the merger, and the relative SFRs measured in these simulations result from merger-driven inflows. And then we find in these models that the relative SFR enhancement by mergers, as well as the duration of these bursts, are rather unchanged compared to galaxies with lower gas fraction.

The case of gas-rich galaxies is then more difficult to study and different assumptions will lead to largely different evolution. However, when one disentangles what drives the star formation activity in each model, one can likely conclude that {\it (i)} merger-driven gas inflows do not trigger more efficient starbursts in gas-rich merging galaxies compared to local galaxies, but {\it (ii)} other processes like internal clump formation, which itself can be provoked by a galaxy interaction, can trigger stronger and longer activities of star formation. We nevertheless note from Appendix~A that such gas clumps  can also occur without mergers/interactions being needed, then triggering comparably high starbursts in non-interacting systems. Internal instabilities may then trigger star formation in all kinds of galaxies and not necessarily enhance the star formation activity in the interacting ones.


\section{Discussion}\label{others}

The role played by interactions and mergers in affecting the evolution of galaxies has been widely studied in the last decades by means of numerical simulations. 
 The first numerical models of interacting galaxies in which all the dynamical components (bulge, disk and massive dark halo) were taken into account by means of fully N-body simulations (with $N\sim 10^4$) were developed by \citet{barnes88, barnes92} and \citet{hs92}. In the same years,  the first attempt to incorporate the dynamics of the interstellar medium and star formation recipes into models of interactions between disk galaxies  \citep{mrb92} showed that the global SFR in merging galaxies could be enhanced by an order of magnitude with respect to pre-interaction levels, for several hundred Myrs, with star formation mostly occurring in the galaxy central regions.\\
These results were confirmed and extended by  \citet{mih94a} and  \citet{mih96}, who incorporated star formation algorithms in a Tree-SPH code  \citep{mih94b} in order to investigate the triggering of starbursts in merging disk galaxies. Their main findings showed that the properties of the merger-driven starburst depend on the morphology of the interacting systems, being galaxies without bulges more prone to disk instabilities in the first phases of the interaction, while central bulges tend to stabilize galaxy disks and to postpone gas inflows and the subsequent starburst to the final merging phase. These works also showed that the merger-driven starburts are very short, with a duration of approximately 50 Myr, but not rare (the SFRs of interacting systems being typically two orders of magnitude greater than that of the two corresponding isolated galaxies). While the simulations discussed in this paper and in DM07 present the same dichotomy in the response of early-type and late-type galaxies to interactions-driven instabilities (see, for example, Figs.10 and 13 in DM07), contrary to these works, our findings show that the most intense star formation enhancements (a factor greater than 10 with respect to that of isolated systems) are  \emph{not only short, but also rare}. Lower star formation enhancements with respect to that of  \citet{mih96} were also found by  \citet{spri00}: even adopting a more sophisticated feedback with respect to that employed by  \citet{mih96} and by us, they find a star formation rate at maximum ten times greater than that of isolated systems, so an order of magnitude lower than the ones proposed by   \citet{mih96}. 
The evolution of the star formation rate in the  Mice, studied by means of numerical simulations by \citet{barnes04}, also suggests star formation rates at most ten times greater than  pre-interaction levels. Interestingly, this work compares different star formation prescriptions, showing how the extensions of the star forming regions can be affected when taking into account also  gas shocks in the modeling.\
The different response of the star formation evolution in interacting galaxies with or without bulges was also confirmed by \citet{sprietal05}. In this work, a stability analysis of isolated disk galaxies was performed, showing the difficulty to prevent disk instabilites when a large isothermal gas fraction is adopted. This is in accordance with the evolution of our gas-rich systems, which, in fact, shows that strong disk instabilities can occur unless  gas dissipation is artificially reduced (as is done for gas-rich galaxies in PM-SP simulations).  They also  pointed out the role played by accreting black holes in quenching star formation during the final galaxy coalescence. 

A simple stellar feedback model was implemented by  \citet{cox06}, in their simulations of disk galaxy major mergers. They adopted a feedback model which stores SNe energy within individual gas particles and dissipates it on a time-scale defined by $\tau_{fb}$ and according to an equation-of-state in star forming regions set by a $n$ parameter. The adopted values of  $\tau_{fb}$ and  $n$  determine different absolute star formation rates both for isolated and merging systems, but, for the most plausible feedback prescriptions (i.e. feedback not too low), the choice of different parameters does not strongly affect the relative star formation rates (see Table 3 in that paper). They also pointed out that the adoption of an isothermal equation of state for the gas can overestimate the star formation rates of merging pairs, because of the underestimation of shock heatings in low gas density regions.

In our simulations we did not focus on the dynamics of gas falling-back from tidal tails, or similarly the gas potentially expelled by strong winds and falling-back at later-stages. Such processes could contribute to increase the star formation activity after the merger, but the dynamical timescale in the outer regions is large so this will likely occur only after the merger remnant is well relaxed rather than during the merger itself. Also, these mechanisms would be expected to feed a large-scale disk rather than a dense central concentration, so the associated star formation would last long with a moderate efficiency instead of consisting of a starburst activity. Dedicated studies would be needed to fully constrain the role of the late gas fall-back on merger remnants; in any case the associated star formation is not related to the merger-driven inflow process that we have studied in the present paper.

\section{Summary and comparison to observations}\label{obs}

\subsection{At low redshift}\label{obs1}

We have inferred from the two simulations dataset presented in this paper that, for galaxies with gas content typical of Local or low-redshift galaxies (typically 15\% of gas in the disk and up to 30\%), the starbursts induced by galaxy major mergers have a moderate intensity, star formation rates being rarely enhanced by factors larger than 5 compared to isolated galaxies, even at the peak of the starbursts. This result appears to be robust because it is found in both  numerical methods used (Tree-SPH and PM-SP) and is not largely affected when different assumptions are made for the determination of the local star formation rate.

A large-scale tidal field from dense cosmological structures could further enhance merger-induced starbursts \citep{mart07}, especially at high redshift. This was not accounted for here, but the average effect does not exceed a factor of two. It obviously does not affect at all mergers in the field or low-mass groups. Also, accounting for feedback or not leaves the statistical conclusions about unchanged when our two codes are compared. The Tree-SPH dataset shows that the star formation activity is comparable in mergers and fly-by interactions. We here focused mainly on interactions of galaxies of comparable masses, but \citet{cox07} show that minor interactions are much less efficient than major ones to trigger star formation.

Strong starbursts can still result from galaxy major interactions and mergers. About fifteen percent on the galaxy interactions in our models have a merger-triggered starburst with a relative efficiency higher than 5. To our knowledge, there is no other mechanism susceptible to trigger such intense starbursts in low-redshift disk galaxies, and this agrees well with the vast majority of LIRGs and ULIRGs in the local universe showing signs of interactions or mergers \citep{sanmir96, ducmm97}. More generally, nearly all mergers in our model lead to a significant enhancement of star formation of at least 50\%, and in the majority of the models the star formation rate is at least doubled during the merger: this also agrees well with the fact that observed mergers show a general trend to enhanced star formation \citep{geor00, char01}.

However, merger-induced starbursts have on average a limited efficiency. Strong starbursts are rare among simulated mergers and interactions, and this again is in agreement with observations: observed starbursts at low redshifts are merger-driven, but merging galaxies do not necessarily show high star formation rates \citep{berg03, cheng07}. Independently of the code used and star formation recipe adopted, we find a median value for the maximal relative SFR of about 3. \citet{jogee07} have studied star formation in interacting galaxies in GEMS at low and moderate redshifts (z $\sim$ 0.24 to 0.80) showing that the average   SFR of strongly   distorted interacting/merging massive galaxies is only
   modestly  enhanced with respect to  normal undisturbed
   systems, in agreement with the results from our models. Note that simulations can result in a slightly higher SFR enhancement than observations  because the instant of the maximal SFR is selected in simulated pairs.

\subsection{At high redshift}

At higher redshift, in particularly gas-rich galaxies, the situation is more complex. It appears that merger-induced gas inflows are not more efficient trigger of star formation than they are at low redshift. The star formation activity can be enhanced in much larger proportions, and over much larger durations, if the gas-rich disks become Jeans unstable and form dense gas clumps during the interaction, as is the case in the Tree-SPH simulations. However, comparison with the PM-SP simulations and other simulations presented in the Appendix indicate that this is a rather particular situation. Such clumps do not form in disks that are too stable -- the starbursts have low average efficiencies -- while in disks that are too unstable the star-forming clumps form even without the interaction -- then only a moderate additional starburst can be triggered if an interaction occurs.

Under specific conditions\footnote{Note, however, that cold gas accretion from filaments in the near environment of galaxies  has not been taken into account in this work. We plan to study this aspect in future works, as it can affect the galaxy disk stability, as well as its star formation rate.} where disks are at the stability limit before merger occurs, merger-driven fragmentation can lead to strong merger-driven starbursts. However, there are evidences that disks can become clumpy as a result of their internal evolution only, without mergers/interactions being required for that \citep{elm07, bourn07a} (see also \citet{bourn08}), which can then be another way to trigger star formation without galaxy interactions.

Observationally, the role of mergers in the star formation history at high redshift is still debated (see Introduction). There are however several studies (e.g. Bell et al. 2005) who find that this role is limited. \citet{daddi07c} find long duty cycles for ULIRGs at high redshift, which are longer than the typical duration of merger-induced starbursts in our simulations, suggesting that another mechanism rather than interactions/mergers is the main trigger of starbursts and AGNs there. 
 Large star formation rates could then be triggered by internal evolution, or simply result from the presence of large gas reservoirs around the most actively star-forming galaxies \citep{daddi07c}.

\section{Conclusion}\label{concl}
In this work we have analyzed the relation between galaxy interactions and star formation enhancement by means of a vast number ($\sim$ 1000, in total) of simulations, varying the numerical code adopted as well as the numerical recipes to model star formation. 
 The main results of this study are the following.\\
\begin{itemize}
\item At low redshifts, interactions and mergers, in general, produce moderate enhancements in the star formation rate of the pairs, while strong starbursts are rare. This result does not depend either on the numerical code adopted, or on the recipes used to account for star formation. For star formation prescriptions compatible with observations ($\Sigma_{SFR} \propto {\Sigma_{gas}}^{1.5}$ or $\Sigma_{SFR} \propto \Sigma_{gas}\Omega$), the majority of the encounters ($\sim$ 85$\%$) leads to a maximal star formation enhancement less than a factor five. As discussed in Sect.\ref{obs1}, these results are in good agreement with several observational studies \citep{berg03, cheng07, jogee07}.
\item The duration of the moderate starbursts is generally smaller than 500 Myr for Tree-SPH models. Even if PM-SP simulations have a somewhat different statistical distribution, both models show that no more than 15~$\%$ of interaction-induced starbursts have a duration greater than 500 Myr. This would suggest that another mechanism has a major role in triggering the activity of high-redshift ULIRGs, which has a longer duty cycle \citep{daddi07a}.
\item Moving to higher redhifts (i.e. essentially increasing the gas mass fraction), interacting systems have absolute SFRs greater than local interacting pairs.
\item Inflow-induced starbursts during interactions and mergers are neither stronger\footnote{The SFR being normalized to that of isolated galaxies.} nor longer than their local counterparts. In turn, Jeans instability in gas-rich disks can cause the formation of massive clumps, as observed \citep{elm07, bourn07a, bourn08}. The disk fragmentation can appear either in interacting either in isolated systems and the induced fragmentation-driven starbursts are characterized by high star formation enhancements as well as long bursts duration, but then interactions and mergers are not necessarily required to trigger this kind of activity.
\end{itemize}

Additional numerical work is required to firmly establish the nature  
of high-$z$ ULIRGs and  understand the role of internal factors  
compared to interactions and mergers in the triggering of the star  
formation activity and/or AGN. Nevertheless, this work presents  
strong results in terms of frequency and duration of starbursts  
episodes triggered by galaxy interactions. Strong starbursts can occur in some major mergers, and this is consistent with submillimiter galaxies being mostly major mergers  \citep[e.g.][]{tacconi08}, but, on average, the classical merger- 
driven gas inflows do not seem to be an efficient process, and other  
mechanisms could have an important role too, like the fragmentation  
of gas-rich disks, or the presence of large gas reservoirs as  
recently suggested by \citet{daddi07c}. These conclusions hold well  
with observations showing that a large number of actively star- 
forming galaxies are massive disks rather than mergers \citep 
{genzel06, FS06, daddi07a, shapiro08}. \citet{elm07} have shown that resolved  
star-forming galaxies in the Hubble Ultra-Deep field at redshift 1  
and above are largely dominated by primordial disks. These high- 
redshift disks outnumber merger candidates; they form stars at high  
rates because they are massive and gas-rich: pervasive clumpy  
structures attest of the high gas content of these massive star- 
forming disks. These results certainly also help to disentangle the  
nature of massive star formation at high redshift, in the light of  
the limited role of mergers in systematically inducing extreme star formation rates, as suggested by our models.

\begin{acknowledgements}

The authors are  grateful to M. Lehnert, for a careful reading of the paper and for all his suggestions, and to Y. Revaz for providing the python parallelized pNbody package (see  \emph{http://obswww.unige.ch/$\sim$revaz/pNbody/}), used for making galaxy maps. \\ 
We wish to thank the referee for his comments, which helped in improving the contents of this paper. \\

This research used  the computational resources of IDRIS and CEA/CCRT and those available within the framework of the Horizon Project (see \emph{http://www.projet-horizon.fr/}).
\end{acknowledgements}

\begin{appendix}
\section{Tree-SPH and PM-SP codes: some validity tests} \label{tests}
\subsection{Tree-SPH}
The main parts of the Tree-SPH code used to run most of the simulations presented in this paper has been described in  \citet{benoit02}. In particular,  we employed the same evaluation of the gravitational forces, the same implementation of the SPH technique and of the star formation modeling as the one described in the above cited paper. Some validity tests of this code have been presented in  \citet{benoit02}, in particular that concerning the collapse of  an initially static, isothermal sphere of self-gravitating gas --a  standard validation test for this type of codes, see  \citet{evrard88} and also  \citet{hk89, thack00, spri01}). 
Here we present some other tests, about the dependency of the star formation rates on the numerical parameters adopted (integration time-step, number of neighbors $N_s$ used in evaluating SPH quantities, value of the gravitational smoothing length $ \epsilon$, and number of total particles in the systems). In particular, we have chosen one of the coplanar merger simulations described in Sect.\ref{increasing} (the one with id=gSb+gSb+09ret00) and we have performed four additional runs, changing:
\begin{itemize}
\item the value of the integration time-step, taken as one half of that of the reference case, i.e. $ \Delta t= 25Myr$;   
\item the value of the number of neighbors $ N_s$ adopted to evaluate the SPH quantities, performing one run with $ N_s=30$, instead of $ N_s= 15$ ;
\item the value of the gravitational smoothing length, assuming $\epsilon=350pc$ instead of $\epsilon=280pc$ ;
\item the total number of particles of the simulation, performing a high-resolution run with a total number of particles equal to $N_{tot}=960000$, i.e. four times higher than the reference case value.
\end{itemize}
The results of this study are presented in   Fig.\ref{fredtest}, left panel, where the relative SFR for the reference case (black curve) and for the additional tests are presented. As it can be seen, the two peaks of the SFR (corresponding to the time of pericenter passage and to the merging phase) are only slightly affected by the choice of different numerical parameters, being the differences in the relative SFR values of $\sim 10\%$ at the first passage, and of  $\sim 20\%$ at most in the final coalescence phase.
 
\subsection{PM-SP}

\begin{figure}
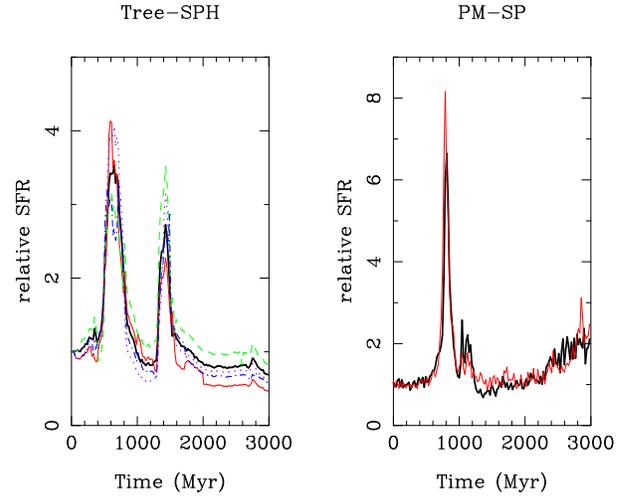

 \begin{minipage}[b]{4.2cm}
   \centering
  \includegraphics[width=6.5cm,angle=270]{mytests_smooth.ps}
 \end{minipage}
 \begin{minipage}[b]{4.2cm}
   \centering
  \includegraphics[width=6.5cm,angle=270]{test_pmsp.ps}
 \end{minipage}
\caption{Left panel: validity tests for the Tree-SPH simulations. The relative SFR evolution during an interaction is shown, either when using the standard values described in Sect.\ref{numerical1} (solid thick black curve), or an integration time-step half of that of the reference case (solid thin red curve), or a gravitational smoothing length 1.4 times higher (dashed green curve), or a number of neighbors $N_s$ two times higher than that of the reference case (dot-dashed blue curve), or a number of total particles in the system four times higher than the reference value (dotted violet curve). Right panel: validity test for the PM-SP simulations. The relative SFR evolution during an interaction is shown, when using either the standard values described in \ref{numerical2} (solid thick black curve) or an integration time-step half of that of the reference case (solid thin red curve). \label{fredtest} }
\end{figure}

Some validity tests have been run also for the PM-SP simulations. In particular, we have analyzed the dependency of the relative SFR on the integration time-step, choosing six representative mergers, which have a maximum relative SFR equal to 4.66 on average. Dividing the initially adopted time step $ \Delta t=1 Myr$ by two, we have checked that the evolution of the relative SFR is the same (cfr. Fig.\ref{fredtest}, right panel), with a maximum SFR which can be at most $ \sim 20\%$ greater or smaller than the reference value, but with no systematic effects with respect to it, which seems to be a sampling effect rather than a physical difference.

\end{appendix}

\begin{appendix}
\section{Some evolutions of absolute star formation rates during interactions and mergers}\label{absolutesfr}

\begin{figure}
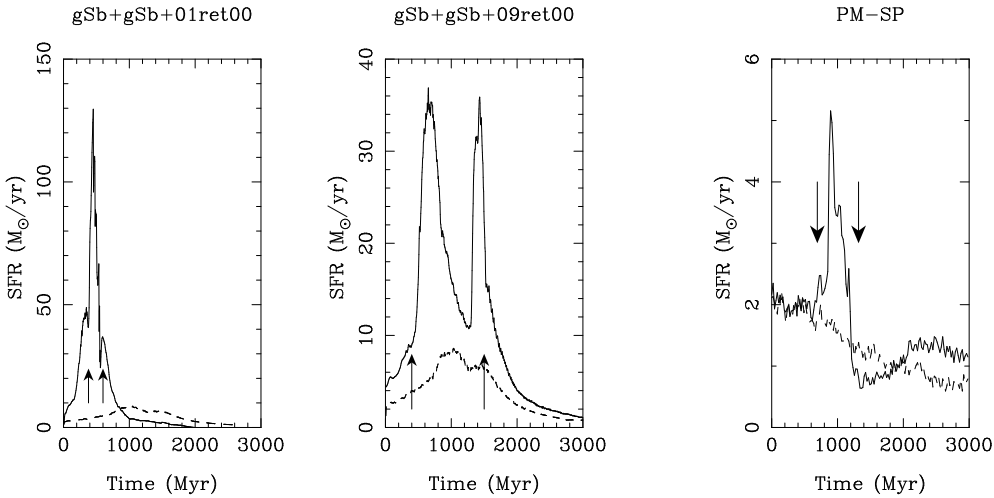

 \begin{minipage}[b]{4.2cm}
   \centering
  \includegraphics[width=6.5cm,angle=270]{absgSbgSb01ret00+.ps}
 \end{minipage}
\begin{minipage}[b]{4.2cm}
   \centering
  \includegraphics[width=6.5cm,angle=270]{absgSbgSb09ret00+.ps}
 \end{minipage}
\caption{Absolute star formation rate, versus time, for some coplanar mergers involving two gas-rich gSb+ galaxies (solid lines). The absolute star formation rate of the isolated gSb+ galaxy is also shown (dotted line). In each panel, the black arrows indicate, respectively, the first pericenter passage between the two galaxies and the merger epoch. See Fig.\ref{sfr+} for the evolution with time of the relative SFR for these two encounters.\label{abssfr+} }
\end{figure}

\begin{figure}
 \begin{minipage}[b]{4.2cm}
   \centering
  \includegraphics[width=6.5cm,angle=270]{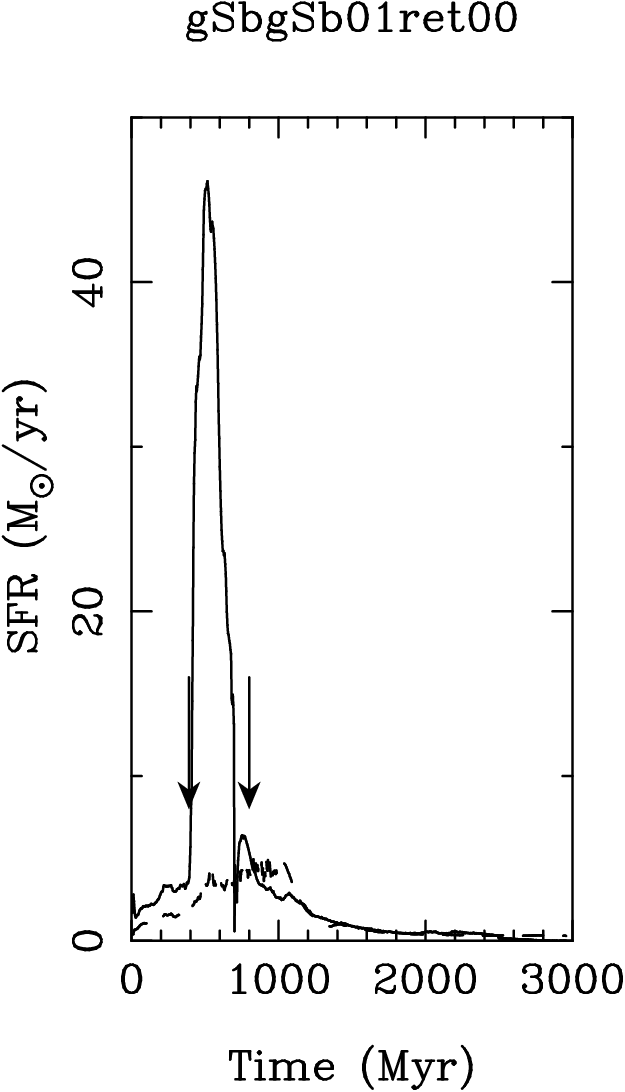}
 \end{minipage}
\begin{minipage}[b]{4.2cm}
   \centering
  \includegraphics[width=6.5cm,angle=270]{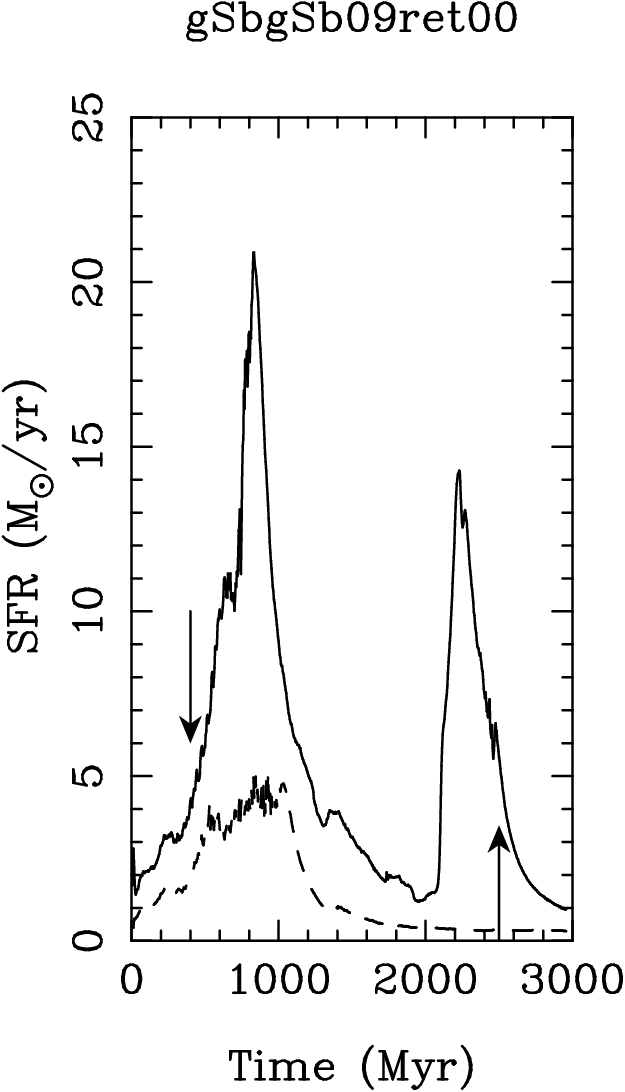}
 \end{minipage}
\caption{Absolute star formation rate, versus time, for some coplanar mergers involving two local gSb galaxies (solid lines). The absolute star formation rate of the isolated gSb galaxy is also shown (dotted line). In each panel, the black arrows indicate, respectively, the first pericenter passage between the two galaxies and the merger epoch. See Fig.\ref{sfrno+} for the evolution with time of the relative SFR for these two encounters.\label{abssfr} }
\end{figure}

\begin{figure}
 \begin{minipage}[b]{4.2cm}
   \centering
  \includegraphics[width=6.5cm,angle=270]{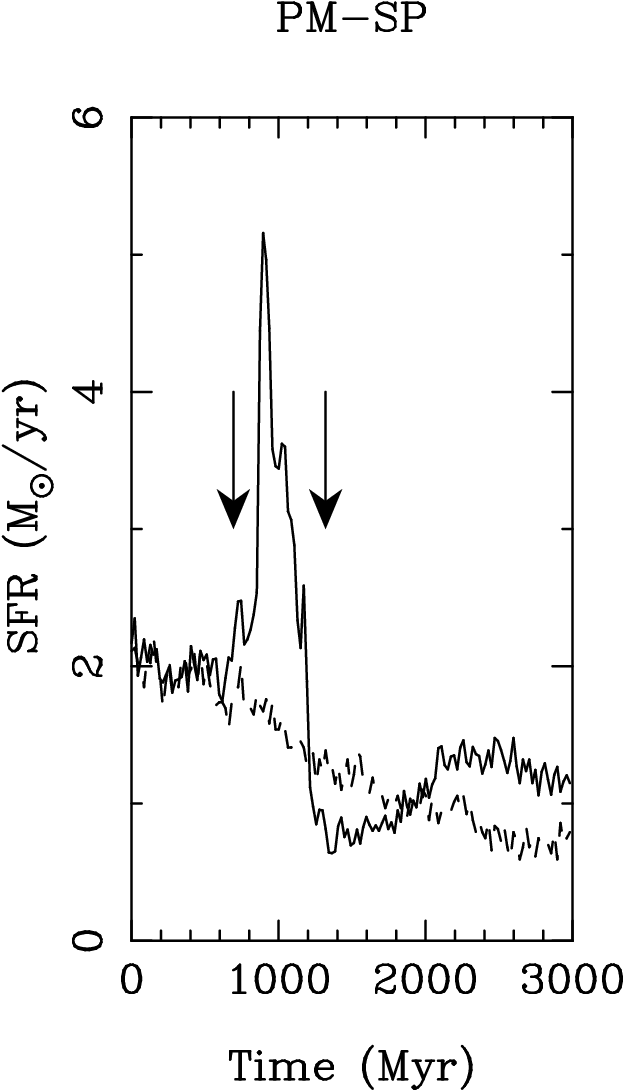}
 \end{minipage}
\caption{Absolute star formation rate, versus time, for  a PM-SP merger (solid line). The absolute star formation rate of the isolated gSb galaxy is also shown (dashed line).  The black arrows indicate, respectively, the first pericenter passage between the two galaxies and the merger epoch.\label{fredabssfr} }
\end{figure}

In order to separate the contribution of tidal effects from secular evolution in determining the star formation history of the interacting pairs, all the star formation rates presented in the paper have been shown relative to that of the two corresponding isolated galaxies. Here we want to show the evolution of the absolute star formation rates during some encounters, in order to give some indications about the absolute rates involved.\\
In Figs. \ref{abssfr+} and  \ref{abssfr} this is done for some coplanar encounters involving two gas-rich  spirals and two local Sb spirals, respectively (cfr Figs. \ref{sfr+} and  \ref{sfrno+} for the evolution of their relative SFRs). In the case of the gSb+gSb+01ret00 interaction, the merging phase occurs soon after the first pericenter passage and the burst of star formation ($SFR \sim 130 M_{\odot}/yr$) takes place between this two phases. At the same epoch ($t \sim 450 Myr $), the isolated gSb+ galaxy sustains a SFR of the order of 4 $M_{\odot}/yr$. This leads  to a normalized star formation rate (relative to that of the two gSb+ galaxies involved in the interaction) of about 15 (see left panel in Figs \ref{abssfr+}). After this starburst, the SFR of the merger decreases rapidly, to the consumption of most of the gas available.\\
The evolution of the star formation history of the gSb+gSb+09ret00 interaction is shown in the right panel of Fig.\ref{abssfr+}: in this case the star formation rate of the interacting galaxies begins to increase rapidly after the first pericenter passage, peaking at about 36  $M_{\odot}/yr$  200 Myr after the first encounter; a second burst of the same absolute magnitude occurs when the coalescence of the two galaxies takes place. As discussed in Sect.\ref{increasing}, the amplitude of this second burst of star formation is limited by the strong kinetic feedback due to SNe explosions, which determines a rapid expansion of the gas component.\\
A similar behaviour to that of gSb+ interactions is found for the SFR evolution of local galaxies (Fig. \ref{abssfr}), except for the fact that these systems contain a smaller gas fraction (cfr Table \ref{galpar}) and so, ultimately, they show lower star formation rates during the interaction.\\
Finally, the evolution of the absolute star formation rate of a representative interacting gSb-Sb pair and of the corresponding isolated galaxy in the case of PM-SP simulations is given in Fig.\ref{fredabssfr}.

\end{appendix}

\begin{appendix}
\section{Disk fragmentation and star formation enhancement}\label{app1}

\begin{figure}
  \centering
  \includegraphics[width=8.8cm,angle=0]{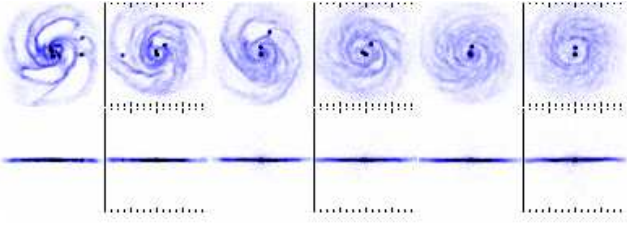}
\caption{Gas maps for the unstable gSb+u1 galaxy, evolved in isolation. From left to right, maps are shown from t=500 Myr to t=3 Gyr,  every 500 Myr. Both xy projection (bottom panels) and xz projection (top panels) are shown. Each box is 40 kpc x 40 kpc in size.\label{un_isogSb+}}
\end{figure}

\begin{figure}
  \centering
  \includegraphics[width=8.8cm,angle=0]{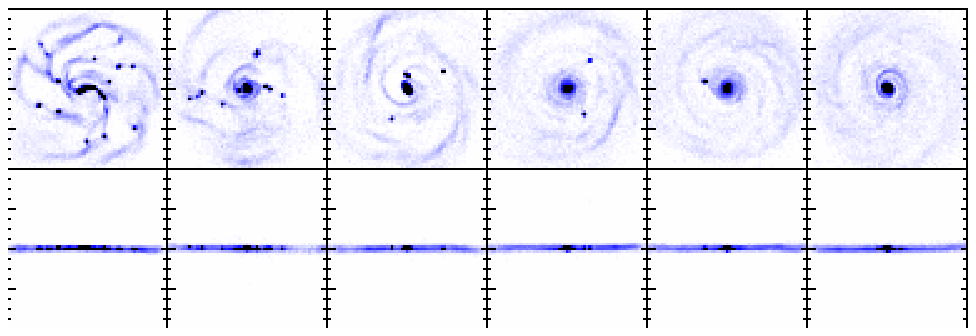}
\caption{Gas maps for the unstable gSb+u2 galaxy, evolved in isolation. From left to right, maps are shown from t=500 Myr to t=3 Gyr,  every 500 Myr. Both xy projection (bottom panels) and xz projection (top panels) are shown. Each box is 40 kpc x 40 kpc in size.\label{plusun_isogSb+}}
\end{figure}

In Sect.\ref{increasing}, we found that  gas-rich spirals in interaction are more prone to disk fragmentation than their local counterparts and that the induced formation of gas clumps is mainly responsible for the star formation enhancement found with respect to isolated systems. To better clarify the role played by disk fragmentation in stimulating star formation, here we want to compare the star formation evolution of the gas-rich gSb+ galaxy (see Fig.\ref{isogSb+} for the gas maps) with that of two gas-rich systems (hereafter called gSb+u1 and gSb+u2),  having the same morphological parameters, but  initial Toomre $Q$ parameters lower than the gSb+ one. In more detail, the Toomre parameter $Q$, equal to unity for the gas component of the gSb+ galaxy, has been lowered to $Q=0.3$ for the gSb+u1 galaxy and to $Q=0.1$ for the gSb+u2 one. \\
The effect of reducing the local disk stability on the galaxy evolution is shown in Fig.\ref{un_isogSb+} and \ref{plusun_isogSb+}: with respect to the stable case (Fig.\ref{isogSb+}), the gas disk fragments in many clumps, particularly evident for the gSb+u2 system. These clumps form, migrate towards the galaxy centers and, in some cases, dissolve under the influence of the tidal field of the disk \citep{bourn07a}.\\

\begin{figure}
  \centering
\includegraphics[width=5.cm,angle=270]{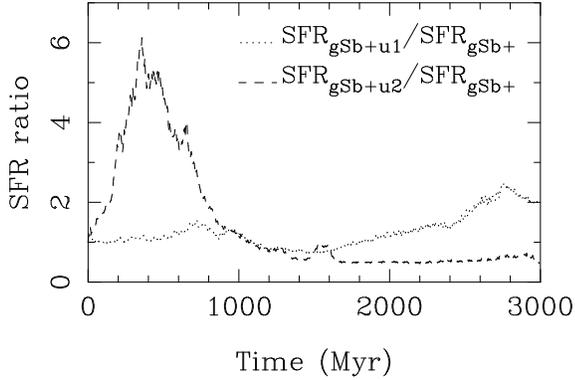}
\caption{Evolution with time of the star formation rates of the  gSb+u1 (dotted line) and gSb+u2 (dashed line) galaxies, normalized to that of the stable gSb+ system.\label{un_sfr}}
\end{figure}

\begin{figure}
  \centering
\includegraphics[width=5.cm,angle=270]{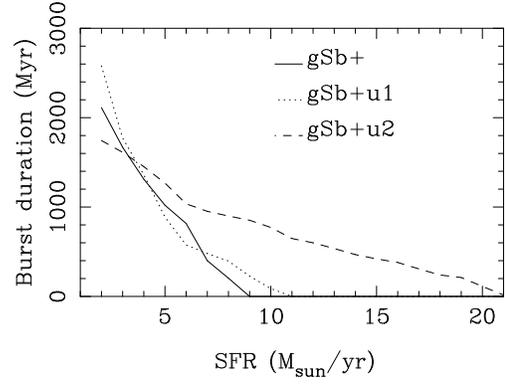}
\caption{Duration of a star formation enhancement greater than the value specified on the x-axis for the stable gSb+ galaxy (solid curve) and for the unstable gSb+u1 (dotted curve) and gSb+u2 (dashed curve) systems.\label{un_time}}
\end{figure}

It is natural to ask if a higher level of gas clumps formation in the disk also results in a different star formation evolution for these isolated systems. As shown in Fig.\ref{un_sfr}, this is indeed the case. Comparing the star formation rate of the unstable gSb+u1 galaxy with that of the more stable gSb+ one, we found that during 3 Gyr of evolution, the unstable disk shows a SFR higher than that of the stable counterpart. This is especially true after t=2 Gyr, when some clumps in the gSb+u1 galaxy are migrating in the central region of the system: in this phase, the ratio between the two star formation rates increases up to a factor of 2.5. Even more striking is the gSb+u2 case, the most unstable one: in this case, a lot of clumps form in the first Gyr of evolution, giving rise to a star formation burst which peaks at 6 times that of the stable gSb+ galaxy. \\
The presence of gas clumps in the disk also affects the duration of the star formation enhancement, in the sense that  the most clumpy is the disk, the longest is the duration of the star formation enhancement. As shown in Fig.\ref{un_time},  in general, the unstable galaxies sustain star formation enhancements greater than a certain threshold (the threshold being varied from 2 $M_{\odot}/yr$ to 21 $M_{\odot}/yr$) for a duration longer than the one of the corresponding stable galaxy. These few examples seem to show that \emph{the most the galaxy fragments under the effects of local disk instabilities, the greatest is the burst enhancement and the longest is the burst duration}.


\end{appendix}

\begin{appendix}
\section{Statistics for Tree-SPH simulations}\label{app2}
 In this Appendix, some more details about the statistical distributions of the intensity, duration and integrated star formation rate of interacting galaxy pairs are given. In particular, three tables are shown. The first one (Table \ref{sfrstat}) presents the statistical distribution of the maximum star formation rates for merger and flybys, whose corresponding histograms are shown in Fig.\ref{istosfr}. The duration of the enhanced star formation rate is given in Table~\ref{timestat}, for two different SFR thresholds. We refer the reader to Sect.\ref{duration} and Fig.\ref{istotime} for a discussion on this. Finally, in Table \ref{isfrstat},  the quartiles of the distribution of the integrated star formation rate, for mergers and flybys, are given. The corresponding histograms are shown in Fig.\ref{istoisfr}.

   \begin{table}
      \caption[]{Quartiles of the distribution of the maximum star formation rate (relative to the isolated case) for mergers and flybys. The first ($P_{25}$), second ($P_{50}$) and third ($P_{75}$) quartiles are given, for all the different disk inclinations. See Fig.\ref{istosfr} for the histograms of the distributions.}
         \label{sfrstat}
     \centering
      \begin{tabular}{lcccccccc}
        \hline\hline
	& \multicolumn{4}{c}{Mergers} & \multicolumn{4}{c}{Flybys}\\
        &  $0^\circ$ & $45^\circ$ & $75^\circ$ & $90^\circ$ &   $0^\circ$ & $45^\circ$ & $75^\circ$ & $90^\circ$ \\
        \hline
        & & & & & & & &  \\
	$P_{25}$ & 2.57 & 2.72 & 2.55 & 2.78 & 2.12 & 1.86 & 1.80 & 1.89\\
	$P_{50}$ & 3.96 & 4.11 & 3.66 & 4.10 & 2.59 & 2.63 & 2.47 & 2.55\\
	$P_{75}$ & 5.95 & 7.65 & 5.64 & 7.00 & 3.62 & 3.30 & 3.12 & 3.32\\
        & & & & & & & &  \\
            \hline

            \hline
         \end{tabular}
   \end{table}

   \begin{table}
      \caption[]{Quartiles of the distribution of the duration (in Myr) of enhanced SFR for the whole sample of interacting galaxies (mergers and flybys). Two thresholds are shown: relative SFR $>$ 2 and relative SFR $>$ 5.  The first ($P_{25}$), second ($P_{50}$) and third ($P_{75}$) quartiles are given, for all the different disk inclinations. See Fig.\ref{istotime} for the  histograms of the distributions.}
         \label{timestat}
     \centering
      \begin{tabular}{lcccccccc}
        \hline\hline
	& \multicolumn{4}{c}{relative SFR $>$ 2} & \multicolumn{4}{c}{relative SFR $>$ 5}\\
        &  $0^\circ$ & $45^\circ$ & $75^\circ$ & $90^\circ$ &   $0^\circ$ & $45^\circ$ & $75^\circ$ & $90^\circ$ \\
        \hline
        & & & & & & & &  \\
	$P_{25}$ & 64. & 32. & 52. &  52. &  36. & 88. & 56.& 36.\\
	$P_{50}$ & 156.& 152. & 152. & 172. & 84. & 239. &96.& 104.\\
	$P_{75}$ & 320.& 364. & 368. & 368. & 164.& 412. & 212.& 147.\\
        & & & & & & & &  \\
            \hline

            \hline
         \end{tabular}
   \end{table}

   \begin{table}
      \caption[]{Quartiles of the distribution of the integrated star formation rate (relative to the isolated case) for mergers and flybys. The first ($P_{25}$), second ($P_{50}$) and third ($P_{75}$) quartiles are given, for all the different disk inclinations. See Fig.\ref{istoisfr} for the histograms of the distributions.}
         \label{isfrstat}
     \centering
      \begin{tabular}{lcccccccc}
        \hline\hline
	& \multicolumn{4}{c}{Mergers} & \multicolumn{4}{c}{Flybys}\\
        &  $0^\circ$ & $45^\circ$ & $75^\circ$ & $90^\circ$ &   $0^\circ$ & $45^\circ$ & $75^\circ$ & $90^\circ$ \\
        \hline
        & & & & & & & &  \\
	$P_{25}$ & 1.10 & 1.02 & 1.04 & 1.07 & 1.04 & 0.94 & 0.98 & 0.99\\
	$P_{50}$ & 1.23 & 1.15 & 1.16 & 1.17 & 1.10 & 1.01 & 1.03 & 1.03\\
	$P_{75}$ & 1.47 & 1.30 & 1.29 & 1.31 & 1.23 & 1.11 & 1.11 & 1.13\\
        & & & & & & & &  \\
            \hline

            \hline
         \end{tabular}
   \end{table}

\end{appendix}


\begin{thebibliography}{}

\bibitem[Arp(1966)]{arp}Arp, H. 1966, ApJS, 14, 1
\bibitem[Avila-Reese et al.(1998)]{avila98}Avila-Reese, V., Firmani, C., \& Hernandez, X. 1998, ApJ, 505, 37 
\bibitem[Avila-Reese \& Firmani(2001)]{avi01}Avila-Reese, V., \& Firmani, C. 2001, RevMexAA, 10, 97
\bibitem[Barnes \& Hut(1986)]{bh86}Barnes, J., Hut, P. 1986, Nature, 324, 446
\bibitem[Barnes(1988)]{barnes88}Barnes, J. 1988, ApJ, 331, 699
\bibitem[Barnes(1992)]{barnes92}Barnes, J. 1992, ApJ, 393, 484
\bibitem[Barnes(2004)]{barnes04}Barnes, J. 2004, MNRAS, 350, 798
\bibitem[Bell et al.(2005)]{bell05}Bell, E. F., Papovich, C., Wolf, C. et al. 2005, ApJ, 625, 23
\bibitem[Bergvall et al.(2004)]{berg03}Bergvall, N., Laurikainen, E. \& Aalto, S. 2003, A\&A, 405, 31
\bibitem[Binney \& Tremaine(1987)]{bt1}Binney, J., \& Tremaine, S. 1987, Galactic Dynamics (Princeton: Princeton Univ. Press) 
\bibitem[Boissier et al.(2003)]{bois03}Boissier, S., Prantzos, N., Boselli, A., et al. 2003, MNRAS, 346, 1215
\bibitem[Bournaud \& Combes(2003)]{BC03}Bournaud, F. \& Combes, F. 2003, A\&A,401, 817 
\bibitem[Bournaud et al.(2004)]{bourn04}Bournaud, F., Duc, P.-A., Amram, P., Combes, F., Gach, J.-L. 2004, A\&A, 425, 813
\bibitem[Bournaud et al.(2007a)]{bourn07a}Bournaud, F., Elmegreen, B. G., Elmegreen, D. M. 2007a, ApJ, 670, 237 
\bibitem[Bournaud et al.(2007b)]{bourn07b}Bournaud, F., Jog, C. J. \& Combes, F.  2007b, A\&A, 476, 1179
\bibitem[Bournaud et al.(2008)]{bourn08}Bournaud, F., Daddi, E., Elmegreen, B. G. et al. 2008,  A\&A, in press
\bibitem[Bridge et al.(2007)]{bridge07}Bridge, C. R., Appleton, P. N., Conselice, C. J., et al. 2007, ApJ, 659, 931
\bibitem[Charmandaris et al.(2001)]{char01}Charmandaris, V., Laurent, O., Mirabel, I. F., Gallais, P. 2001, ApSSS, 277, 55
\bibitem[Chemin et al.(2007)]{chemin07}Chemin, L., Carignan, C., Amram, P. 2007, in Proceedings of ``Pathways through an Eclectic Universe'', April 2007, Johan Knapen, Tarry Mahoney  \& Alexandre Vazdekis eds.
\bibitem[Chemin et al.(2008)]{chemin08}Chemin, L., et al. 2008, in preparation
\bibitem[Cole \& Lacey (1996)]{cole96}Cole, S.  \& Lacey, C. 1996, MNRAS, 281, 716
\bibitem[Combes et al.(2006)]{combes06}Combes, F., García-Burillo, S., Braine, J., et al. 2006, A\&A, 460L, 49
\bibitem[Combes(2007)]{combes07}Combes, F. 2007, in Proceedings of "Formation and Evolution of Galaxy Bulges", July 2007, M. Bureau, E. Athanassoula, B. Barbuy eds., astro-ph/0709.0091
\bibitem[Conselice et al.(2003)]{cons03}Conselice, C. J., Chapman, S. C., \& Windhorst, R. A., 2003, ApJL, 596, 5
\bibitem[Cox et al.(2006)]{cox06}Cox, T. J., Jonsson, P., Primack, J., et al. 2006, MNRAS, 373, 1013
\bibitem[Cox et al.(2008)]{cox07}Cox, T. J., Jonsson, P., Somerville, R. S., Primack, J. L., Dekel, A. 2008, MNRAS in press, astro-ph/0709.3511 
\bibitem[Daddi et al.(2007a)]{daddi07a}Daddi, E., Dickinson, M., Morrison, G., et al. 2007a, ApJ, 670, 156
\bibitem[Daddi et al.(2007b)]{daddi07b}Daddi, E., Alexander, D. M., Dickinson, M., et al. 2007b, ApJ, 670, 173
\bibitem[Daddi et al.(2008)]{daddi07c}Daddi, E., Dannerbauer, H., Elbaz, D., et al. 2008, ApJL, 673, 21
\bibitem[de Blok et al.(2001)]{deblok01}de Blok, W. J. G., McGaugh, S. S., \& Rubin, V. C. 2001, AJ, 122, 2396
\bibitem[de Blok et al.(2003)]{deblok03}de Blok, W. J. G., Bosma, A., McGaugh, S. S. 2003, MNRAS, 340, 657
\bibitem[de Grijs(2001a)]{deg01a}de Grijs, R. 2001, A\&G, 42,4
\bibitem[de Grijs(2001b)]{deg01b}de Grijs, R., O'Connell, R. W., Gallagher, J. S. III 2001, AJ, 121, 768
\bibitem[Diemand et al.(2005)]{diemand05}Diemand, J., Zemp, M., Moore, B., Stadel, J.,  \& Carollo, M. 2005, MNRAS, 364, 665
\bibitem[Di Matteo et al.(2007)]{dimatteo07}Di Matteo, P., Combes, F., Melchior, A.-L., \& Semelin, B. 2007, A\&A, 468, 61
\bibitem[Duc et al.(1997)]{ducmm97}Duc,  P.-A., Mirabel, I. F. \& Maza, J. 1997, A\&AS, 124, 533
\bibitem[Duc et al.(2004)]{duc04}Duc, P.-A., Bournaud, F. \& Masset, F. 2004, A\&A, 427, 803
\bibitem[Elbaz \& Cesarsky(2003)]{elbces03}Elbaz, D. \& Cesarsky, C. J. 2003, Science, 300, 270
\bibitem[Elmegreen(1997)]{elmegreen97}Elmegreen, B. G. 1997, RMxAC, 6, 165
\bibitem[Elmegreen(2002)]{elmegreen02}Elmegreen, B. G. 2002, ApJ, 577, 206
\bibitem[Elmegreen et al.(2007)]{elm07}Elmegreen, D. M., Elmegreen, B. G., Ravindranath, S., Coe, D. A. 2007, ApJ, 658, 763 
\bibitem[Evrard(1988)]{evrard88}Evrard, A. E. 1988, MNRAS, 235, 911
\bibitem[Flores \& Primack(1994)]{flores94}Flores, R. A. \& Primack, J. R. 1994, ApJ, 427, L1
\bibitem[F{\"o}rster Schreiber et al.(2006)]{FS06} F{\"o}rster Schreiber, N.~M., et al.\ 2006,  
\apj, 645, 1062
\bibitem[Gao \& Solomon(2004)]{gs04} Gao, Y., \& Solomon, P. M. 2004, ApJ, 606, 271
\bibitem[Gentile et al.(2005)]{gentile05}Gentile, G., Burkert, A., Salucci, P., Klein, U., \& Walter, F. 2005, ApJ, 634, L145
\bibitem[Genzel et al.(2006)]{genzel06} Genzel, R., et al.\ 2006, \nat, 442, 786
\bibitem[Giallongo et al.(1999)]{giallo99}Giallongo, E., Menci, N., Poli, F., D'Odorico, S., \& Fontana, A. 2000, ApJ, 530, L73
\bibitem[Georgakakis et al.(2000)]{geor00}Georgakakis, A., Forbes, D. A., Norris, R. P. 2000, MNRAS, 318, 124
\bibitem[Gingold \& Monaghan(1982)]{gm82}Gingold, R. A., \& Monaghan, J. J. 1982, JCoPh, 46, 429
\bibitem[Hernquist \& Katz(1989)]{hk89}Hernquist, L., \& Katz, N. 1989, ApJS, 70, 419
\bibitem[Hernquist \& Spergel(1992)]{hs92}Hernquist, L., \& Spergel, D. N. 1992, ApJ, 399, L117
\bibitem[Hernquist(1993)]{hern93}Hernquist, L. 1993, ApJS, 86, 389
\bibitem[James(1977)]{james77}James, R. A. 1977, J. Comput. Phys., 25, 71
\bibitem[Jogee et al.(2007)]{jogee07}Jogee, S. et al. 2007, in Proceedings of ``Formation and Evolution of Galaxy Disks'', October 2007,  Jose G. Funes, SJ and Enrico M. Corsini eds, astro-ph/0802.3901
\bibitem[Jogee et al.(2008 in preparation)]{jogee08}Jogee, S. et al. 2008, in preparation
\bibitem[Kapferer et al.(2005)]{kap05}Kapferer, W., Knapp, A., Schindler, S., et al. 2005, A\&A, 438, 87
\bibitem[Katz(1992)]{kat92}Katz N. 1992,  ApJ 391, 502
\bibitem[Kennicutt et al.(1996)]{saas}Kennicutt, R. C., Schweizer, F., Barnes, J. E., Friedli, D., Martinet, L., \& Pfenniger, D. 1998, Galaxies: Interactions and Induced Star Formation (Berlin: Springer)
\bibitem[Kennicutt(1998a)]{ken98a}Kennicutt, R. C., Jr. 1998a, ARA\&A, 36, 189
\bibitem[Kennicutt(1998b)]{ken98b}Kennicutt, R. C., Jr. 1998b, ApJ, 498, 541
\bibitem[Kennicutt et al.(2006)]{ken05}Kennicutt, R. C., Calzetti, D., Walter, F., et al. 2005, AAS, 207, 6314
\bibitem[Kuzio de Naray et al.(2006)]{kuzio06}Kuzio de Naray, R., McGaugh, S. S., de Block, W. J. G. \& Bosma, A., 2006, ApJS, 165, 461
\bibitem[Kuzio de Naray et al.(2008)]{kuzio08}Kuzio de Naray, R., McGaugh, S. S.,  \& de Block, W. J. G., 2008, ApJ, 676, 943
\bibitem[Larson \& Tinsley(1978)]{larson}Larson, R. B.  \& Tinsley, B. M.  1978, ApJ, 219, 46 
\bibitem[Li et al.(2007)]{cheng07}Li, C., Kauffmann, G., Heckman, T. et al. 2007, MNRAS, submitted, astro-ph/0711.3792
\bibitem[Lucy(1977)]{lucy77}Lucy, L. B. 1977, AJ, 82, 1013
\bibitem[Mao et al.(1998)]{maoetal98}Mao, S., Mo, H.J., \& White, S.D.M. 1998, MNRAS, 297, L71
\bibitem[Marchesini et al.(2002)]{marchesini02}Marchesini, D., D'Onghia, E.,  Chincarini, G., Firmani, C., Conconi, P., Molinari, E., \& Zacchei, A. 2002, ApJ, 575, 801
\bibitem[Martig \& Bournaud(2007)]{mart07}Martig, M. \& Bournaud, F. 2007, MNRAS in press, astro-ph/0712.0289
\bibitem[McGaugh et al.(2001)]{mc01}McGaugh, S. S., Rubin, V. C., \& de BVlok, W. J. G. 2001, AJ, 122, 2381
\bibitem[Mihos et al.(1992)]{mrb92}Mihos C., Richstone, D. O., \& Hernquist, L. 1992, ApJ, 400,  153
\bibitem[Mihos \& Hernquist(1994a)]{mih94a}Mihos C., \& Hernquist, L. 1994a, ApJL, 431, 9
\bibitem[Mihos \& Hernquist(1994b)]{mih94b}Mihos C., \& Hernquist, L. 1994b, ApJ, 437, 611
\bibitem[Mihos \& Hernquist(1994c)]{mih94c}Mihos C., \& Hernquist, L. 1994c, ApJL, 425, 13
\bibitem[Mihos \& Hernquist(1996)]{mih96}Mihos C., \& Hernquist, L. 1996, 464, 641
\bibitem[Mihos(2004)]{mihos04}Mihos J. C., 2004, in Mulchaey J. S., Dressler A., Oemler A., eds, Clusters of Galaxies: Probes of Cosmological Structure and Galaxy Evolution. Cambridge Univ. Press, Cambridge, p. 277
\bibitem[Navarro et al.(1996)]{navarro96}Navarro, J. F., Frenk, C. S. \& White, S. 1996, ApJ, 462, 563
\bibitem[Navarro et al.(1997)]{navarro97}Navarro, J. F., Frenk, C. S. \& White, S. 1997, ApJ, 490, 493
\bibitem[Sandage(1961)]{sandage}Sandage, A. 1961, The Hubble Atlas of Galaxies..
\bibitem[Sanders \& Mirabel(1996)]{sanmir96}Sanders \& Mirabel, 1996, ARA\&A, 34, 749
\bibitem[Semelin \& Combes(2002)]{benoit02}Semelin, B. \& Combes, F. 2002, A\&A, 388, 826
\bibitem[Shapiro et al.(2008)]{shapiro08} Shapiro, K.~L., et al.\ 2008, ArXiv e-prints, 802, arXiv:0802.0879
\bibitem[Silk(1997)]{silk97}Silk, J. 1997, ApJ, 481, 703
\bibitem[Springel(2000)]{spri00}Springel, V. 2000, MNRAS, 312, 859
\bibitem[Springel et al.(2001)]{spri01}Springel, V., Yoshida, N., \& White, S. D. M. 2001, New Astron., 6, 79
\bibitem[Springel \& Hernquist(2003)]{spr03}Springel, V., \& Hernquist, L. 2003, MNRAS 339, 312 
\bibitem[Springel et al.(2005)]{sprietal05}Springel, V., Di Matteo, T., \& Hernquist, L. 2005, MNRAS, 361, 776
\bibitem[Steinmetz \& M\"uller(1994)]{ste94}Steinmetz M., \& M\"uller E. 1994, A\&A 281, L97
\bibitem[Struck(2006)]{stru05}Struck, C. 2006, Astrophysics Update 2, 115, astro-ph/0511335
\bibitem[Swaters et al.(2003)]{swaters03}Swaters, R. A., Madore, B. F.,  van den Bosch, F. C.,  \& Balcells, M. 2003, ApJ, 583, 732
\bibitem[Tacconi et al.(2008)]{tacconi08} Tacconi, L.~J., Genzel, R., Smail, I. et al.\
2008, \apj, 680, 246 
\bibitem[Thacker et al.(2000)]{thack00}Thacker, R. J., Tittley, E. R., Pearce, F. R., Couchman, H. M. P., \& Thomas, P. A. 2000, MNRAS, 319, 619
\bibitem[van den Bosch al.(2000)]{van00}van den Bosch., F. C., Robertson, B. E., Dalcanton, J. J., \& de Blok, W. J. G. 2000, AJ, 119, 1579
\bibitem[van den Bosch \& Swaters(2001)]{van01}van den Bosch., F. C., \& Swaters, R. A. 2001, MNRAS, 325, 1017
\bibitem[Weedman et al.(1981)]{weed81}Weedman, D. W., Feldman, F. R., Balzano, V. A., Ramsey, L. W., Sramek, R. A., Wuu, C.-C. 1981, ApJ, 248, 105
\bibitem[Wang et al.(2004)]{wang04}Wang, Z., Fazio, G. G., Ashby, M. L. N., et al. 2004, ApJS, 154, 193

\bibitem[Wong \& Blitz(2002)]{wb02}Wong, T., \& Blitz, L. 2002, ApJ, 569, 15

\end{thebibliography}
\end{document}